%% file: iclr2026_conference.tex
\documentclass{article} 
\usepackage{iclr2026_conference,times}
\iclrfinalcopy
\PassOptionsToPackage{numbers}{natbib}
\usepackage{CJK}
\usepackage[utf8]{inputenc} 
\usepackage[T1]{fontenc}    
\usepackage{hyperref} 
\usepackage{url}
\usepackage{booktabs} 
\usepackage{amsfonts} 
\usepackage{nicefrac} 
\usepackage{microtype}
\usepackage{multirow}
\usepackage[table,xcdraw]{xcolor}
\usepackage{colortbl}
\usepackage{subfigure}

\usepackage{ulem}
\usepackage{amsmath}
\usepackage{graphicx}  
\usepackage{wrapfig}
\usepackage{enumitem}
\usepackage{xspace}
\usepackage{mdframed}
\usepackage{amsmath} 

\newcommand{\methodnameshort}{NextQuill\xspace}

\title{\includegraphics[height=1em]{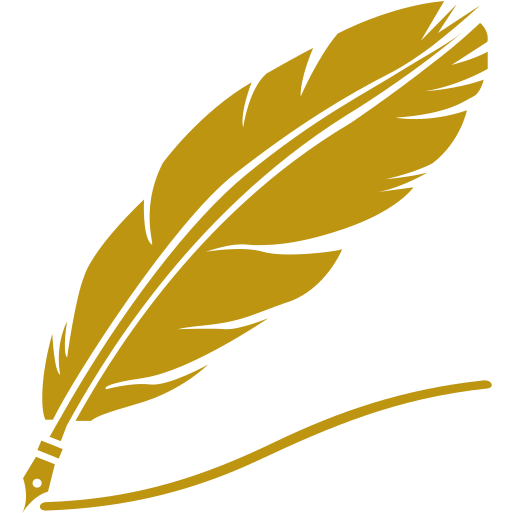}\methodnameshort: Causal Preference Modeling for Enhancing LLM Personalization
}
\input{math_commands.tex}

\usepackage{hyperref}
\usepackage{url}
\usepackage{algorithm}       
\usepackage{amsmath}         
\usepackage{amssymb}         
\usepackage[noend]{algpseudocode} 

\usepackage{tcolorbox}    
\tcbuselibrary{skins, breakable} 
\usepackage{xcolor}        
\usepackage{pifont}        
\usepackage{amsmath, amssymb} 

\newtcolorbox[auto counter, number within=section]{promptbox}[2][]{
  colframe=orange!60!black,
  colback=orange!5,
  colbacktitle=orange!70!black!80,
  coltitle=white,
  left=0mm,
  right=0mm,
  boxrule=1pt,
  borderline={0.8pt}{0pt}{orange!50!black},
  top=1mm,
  bottom=1mm,
  arc=1mm,
  breakable,
  title={\textbf{Prompt~\thetcbcounter: #2}},
  #1
}

\title{\includegraphics[height=1em]{logo.png}\textit{NextQuill}: Causal Preference Modeling for Enhancing LLM Personalization}

\author{Xiaoyan Zhao\textsuperscript{1}\thanks{Equal contribution.}
~~ Juntao You\textsuperscript{2}\footnotemark[1] ~~ \textbf{Yang Zhang\textsuperscript{3}}\thanks{Corresponding author. Email: zyang1580@gmail.com, wenjiewang96@gmail.com.} ~~ \textbf{Wenjie Wang\textsuperscript{2}}\footnotemark[2] ~~ \textbf{Hong Cheng}\textsuperscript{1} \\ 
    \textbf{Fuli Feng\textsuperscript{2}} ~~ 
    \textbf{See-Kiong Ng\textsuperscript{3}}~~
    \textbf{Tat-Seng Chua\textsuperscript{3}} \\
  \textsuperscript{1}The Chinese University of Hong Kong ~~
  \textsuperscript{2}University of Science and Technology of China\\
  \textsuperscript{3}National University of Singapore 
  \\
\texttt{xzhao@se.cuhk.edu.hk}, ~~ \texttt{ustcyjt@mail.ustc.edu.cn}
}

\begin{document}

\maketitle

\begin{abstract}

Personalizing large language models (LLMs) is increasingly important as they are progressively integrated into real-world applications to support users’ daily lives. However, existing approaches often fail to distinguish which components of response predictions by model and ground-truth response in training data truly reflect user preferences, resulting in shallow personalization alignment. In this paper, we introduce NextQuill, a novel LLM personalization alignment framework grounded in causal preference modeling. We approach personalization from a causal perspective, recognizing that model-predicted responses (model side) and user-written ground-truth responses (data side) are both outcomes shaped by user history (characteristics) and other context factors. To better capture user preferences, we define causal preference effects as the causal effect of the  user history/characteristics on outcomes from the model/data side. Building on this foundation, NextQuill introduces two complementary alignment strategies: (1) aligning model-side causal preference effects (on predictions) with those of ground-truth data, rather than indiscriminately aligning all predictions, and (2) emphasizing learning the preference-driven ground-truth tokens, identified via data-side causal preference effects, rather than treating all tokens equally. As such, NextQuill shifts the alignment process toward learning from causal preference effects, facilitating more effective and personalized LLM adaptation. Experiments on multiple personalization benchmarks demonstrate that NextQuill substantially improves personalization quality. Code is available at 
\url{https://github.com/juntaoyou/NextQuill}.

\end{abstract}

\input{1_intro}

\section{Problem Formulation}
 {We study the task of personalized text generation, which aims to enhance an LLM $\mathcal{M}_\theta$ by incorporating user-specific information to generate responses tailored to individual users.}
Given a user query $x$, which specifies the item or objective the user focuses on, a general LLM produces a response $\hat{y} = \mathcal{M}_\theta(x)$. 
To enable personalization, we assume each user is associated with historical text information $h$ that reflects their preferences when having the query. 
{The ground-truth response is denoted by $y$, representing text written by the user.} Then, the triplet $(x, h, y)$ forms a data sample for the user, and the collection of such samples constitutes the user dataset $\sD$, having $(x, h, y) \in \sD$.
Our goal is to develop a model that utilizes both the input query $x$ and the associated profile $h$ to generate a personalized response $\hat{y} = \mathcal{M}_\theta(x, h)$. Specifically, we aim to: (1) improve the preference modeling ability of $\mathcal{M}_\theta$ by effectively incorporating personalized information from user history $h$;
(2) better align the generated response $\hat{y}$ with the personalized information embedded in the expected response $y$.
Ultimately, improve the $\mathcal{M}_\theta$ to generate personalized responses $\hat{y}$.

\input{3_method}

\section{Experiments}
\label{experiment_label}
\subsection{Experimental Settings}\label{sec:expsetting}

\begin{table}[t]
\centering
\large
\caption{Main results of our NextQuill method on four benchmark datasets. \textbf{Bold} numbers denote the best performance, while \underline{underlined} numbers denote the second best. }
\resizebox{\textwidth}{!}{
\fontsize{18}{30}\selectfont
\begin{tabular}{lcccccccccc}
\toprule
\multirow{2}{*}{\textbf{Datasets}} & \textbf{Methods ($\rightarrow$)} 
    & \textbf{Base} &\multicolumn{4}{c}{\textbf{Retrieval-based Methods}} 
    & \multicolumn{4}{c}{\textbf{PEFT-based Methods}} \\ 
\cmidrule(lr){3-3} \cmidrule(lr){4-7} \cmidrule(lr){8-11}
& \textbf{Metrics ($\downarrow$)} & \textbf{Qwen} & \textbf{Contriever} & \textbf{LatestK} & \textbf{LLM-TRSR} & \textbf{CoS} 
  & \textbf{SFT} & \textbf{OPPU} & \textbf{ContextSFT} & \textbf{NextQuill} \\
\midrule
& \textbf{ ROUGE-1}         & { 0.0519} & 0.0507& { 0.0752}  & { 0.0465}   & { 0.0455} & { 0.0717} & { 0.1502} & \uline{0.1661}& { \textbf{0.2318}}                    \\
 & \textbf{ ROUGE-L}         & { 0.0267} & 0.0275& { 0.0406}  & { 0.0269}   & { 0.0278} & { 0.0333} & { 0.0750} & \uline{0.0836}& { \textbf{0.1270}}                    \\
 & \textbf{ METEOR}          & { 0.0639} & 0.0599& { 0.0851}  & { 0.0470}   & 0.0337                        & { 0.0884} & { 0.1013} & \uline{0.2158}& { \textbf{0.2580}}                    \\
 & \textbf{ BLEU}            & { 0.0591} & 0.2064& { 1.3314}  & { 0.5733}   & { 0.3095} & { 0.1849} & { 0.6935} & \uline{2.1214}& { \textbf{3.5718}}                    \\
 \multirow{-5}{*}{\parbox[c]{2cm}{\centering \textbf{Book\\Review}}}   
 & \textbf{ BERTScore}    & { 0.7385} & 0.7414& { 0.7235}  & { 0.7997}   & { 0.7141} & { 0.7841} & { \uline{0.8020}} & { 0.8013}     & { \textbf{0.8182}}                    \\
\midrule
 & \textbf{ ROUGE-1}         & { 0.0470} & 0.0516& { 0.0527}  & { 0.0924}   & { 0.0145} & { 0.0599} & { 0.1231} & \uline{0.1573}& { \textbf{0.2015}}                    \\
 & \textbf{ ROUGE-L}         & { 0.0255} & 0.0290& { 0.0292}  & { 0.0481}   & { 0.0085} & { 0.0301} & { 0.0648} & \uline{0.0753}& \textbf{0.1041}             \\
 & \textbf{ METEOR}          & { 0.0575} & 0.0580& { 0.0582}  & { 0.0918}   & 0.0107                        & { 0.0697} & { 0.0796} & \uline{0.1718}& \textbf{0.1918}               \\
 & \textbf{ BLEU}            & { 0.0402} & 0.2454& { 0.3178}  & { 1.2885}   & { 0.0091} & { 0.1422} & { 0.2555} & \uline{1.7151}& \textbf{2.3845}               \\
\multirow{-5}{*}{\parbox[c]{2cm}{\centering \textbf{Movie \\Review}}}
 & \textbf{ BERTScore}    & { 0.7354} & 0.7411& { 0.7395}  & { 0.7373}   & { 0.6731} & { 0.7445} & { \uline{0.7970}} & { 0.7943}     & \textbf{0.8064}               \\
       \midrule
 & \textbf{ ROUGE-1}         & { 0.0438} & 0.0501& { 0.0513}  & { 0.1074}   & { 0.0327} & { 0.0692} & { 0.1405} & \uline{0.1505}& \textbf{0.1976}               \\
 & \textbf{ ROUGE-L}         & { 0.0231} & 0.0271& { 0.0282}  & { 0.0566}   & { 0.0176} & { 0.0324} & { \uline{0.0714}} & \uline{0.0714}& \textbf{0.0998}               \\
 & \textbf{ METEOR}          & { 0.0517} & 0.0545& { 0.0560}  & { 0.0976}   & 0.0234                        & { 0.0768} & { 0.0848} & \uline{0.1583}& { \textbf{0.1805}}                    \\
 & \textbf{ BLEU}            & { 0.0330} & 0.2386& { 0.2688}  & { 1.2250}   & { 0.0780} & { 0.1220} & { 0.3090} & \uline{1.3487}& { \textbf{1.9676}}                    \\
 \multirow{-5}{*}{\parbox[c]{2cm}{\centering \textbf{CD\\Review}}}
 & \textbf{ BERTScore}    & { 0.7311} & 0.7380& { 0.7375}  & { 0.7479}   & { 0.6900} & { 0.7457} & { \uline{0.7988}} & { 0.7901}     & { \textbf{0.8060}}                    \\
           \midrule
 & \textbf{ ROUGE-1}         & { 0.0684} & 0.0536& { 0.0498}  & { 0.0618}   & { 0.0523} & { 0.0345} & { \uline{0.1229}} & { 0.0934}     & { \textbf{0.1510}}                    \\
 & \textbf{ ROUGE-L}         & { 0.0353} & 0.0286& { 0.0263}  & { 0.0362}   & { 0.0276} & { 0.0198} & { \uline{0.0621}} & { 0.0444}     & { \textbf{0.0729}}                    \\
 & \textbf{ METEOR}          & { 0.0520} & 0.0592& { 0.0568}  & { 0.0574}   & { 0.0672} & { 0.0451} & { 0.0853} & { \uline{0.1219}}     & { \textbf{0.1542}}                    \\
 & \textbf{ BLEU}            & { 0.1233} & 0.0786& { 0.0678}  & { \uline{0.2888}}   & { 0.0460} & { 0.0247} & { 0.2723} & { 0.1981}     & { \textbf{0.5799}}                    \\
 \multirow{-5}{*}{\parbox[c]{2cm}{\centering \textbf{Topic\\Writing}}} 
 & \textbf{ BERTScore}    & { 0.7318} & 0.7464& { 0.7440}  & { 0.7402}   & { 0.7516} & { 0.7424} & { \textbf{0.7990}} & { 0.7827}     & { \uline{0.7957}}                    \\
 \bottomrule                        \end{tabular}
}
\label{tab: main}
\end{table}

\paragraph{Datasets.} 
We conduct experiments on three benchmark datasets from Amazon~\citep{amazon}\footnote{\url{https://amazon-reviews-2023.github.io/}}, including: \textit{Book Review}, \textit{Movie Review}, and \textit{CD Review}, which have been widely used in recent studies on personalized text generation~\citep{DPL}. The task involves generating personalized product reviews that reflect the user’s preferences. Each user–item interaction contains the item title, item description, user rating, and review title, providing rich signals for personalization.
In addition, we include the \textit{Topic Writing} dataset~\citep{kumar2024longlamp}, where the task is to generate a personalized long-form Reddit post on a given topic based on a user-written summary. For each user, we construct a temporally ordered sequence of past interactions as user history. To balance sufficient historical context with the computational cost of long-sequence processing, we follow prior work~\citep{DPL} and cap the input length at 4096 tokens, matching the maximum context window of our backbone LLM.
\paragraph{Baselines.} 
We compare our method against a diverse set of strong baselines spanning three categories.   \textit{{(1) Base Model:}} \textbf{{Qwen}}~\citep{qwen2.5} is a strong LLM backbone without any personalization.
\textit{{(2) Retrieval-based Methods:}} \textbf{{Contriever}}~\citep{lei2023unsupervised} is a widely used dense retriever model, retrieve the most relevant past user interactions for instruction.  
\textbf{{LatestK}}~\citep{liu2024llms+} selects the latest $K$ user interactions based on timestamp, assuming recency reflects more preference.  
\textbf{{CoS}}~\citep{cos} amplifies the influence of context through controlling the activation during decoding.  
\textbf{{LLM-TRSR}}~\citep{llm-trsr} uses recurrent summarization to compress user history into a structured representation.
\textit{{(3) PEFT-based Methods:}}
\textbf{{SFT}}~\citep{lora} is the Standard supervised fine-tuning with the task-specific dataset. 
\textbf{{ContextSFT}}~\citep{salemi2024optimization} is a supervised fine-tuning method that directly trains the LLM using history-augmented information.  
\textbf{{OPPU}}~\citep{oppu} is a PEFT method that learns lightweight user-specific adapters to encode preference signals. Notably, OPPU requires training a separate model for each individual user, unlike all other baselines. This design provides an upper bound for personalization performance but incurs prohibitive computational costs.

\paragraph{Evaluation Metrics.} 

To evaluate the quality of generated reviews, we adopt a comprehensive set of metrics widely used in text generation. We report \textit{ROUGE}~\citep{rouge}, \textit{METEOR}~\citep{meteor}, and \textit{BLEU}~\citep{sacrebleu} to measure lexical and semantic overlap with ground-truth reviews. In addition, we employ \textit{BERTScore}~\citep{zhangbertscore} to assess semantic similarity in the embedding space. Finally, following recent advances in automatic evaluation, we employ an \textit{LLM-as-a-judge} approach~\citep{qwen2.5}, where \texttt{Qwen2.5-72B-Instruct-AWQ} is prompted to provide holistic judgments (LLMScore) on both content relevance and stylistic alignment. The detailed results of this evaluation are presented in Appendix~\ref{appx:llm_eval}, offering a more comprehensive perspective on personalized text generation quality.

\paragraph{Implementation Details.} 
We use Qwen2.5-3B~\citep{qwen2.5} as the backbone LLM for all methods.  For PEFT Methods, we adopt low-rank adaptation (LoRA) \citep{lora} to train our models. We use the AdamW~\citep{adamw} optimizer with a learning rate of \(5 \times 10^{-6}\), a weight decay of 0.025, and a dropout rate of 0.05. Besides, we employ Deepspeed~\citep{deepspeed} for acceleration and ZeRO~\citep{ZERO} for optimization. The train epochs is set to 5. For our framework, the threshold $\delta$ is set to 0.05. The parameter $\alpha$, which controls the weight of the personalized loss, is tuned in the range \{0.01, 0.05, 0.1\}. The high weight $\lambda$ is searched in \{0.9, 0.8, 0.7\}, while the low weight $\epsilon$ is set to 0.1.  

\subsection{Overall Performance}
We compare \methodnameshort with a wide range of baselines and report the results in Table~\ref{tab: main}.
Across all datasets and evaluation metrics, \methodnameshort consistently achieves the best performance, demonstrating the effectiveness of our causal preference modeling framework in capturing and leveraging user-specific information.
Notably, methods that incorporate user information (e.g., CoS, LLM-TRSR) significantly outperform those that do not (e.g., Qwen), confirming the critical role of user history in improving generation quality.
Moreover, PEFT-based methods generally surpass retrieval-only approaches, suggesting that directly optimizing user-conditioned representations is more effective than simply injecting retrieved content.
These findings validate our core design: modeling both the \textit{data-side causal path} for ground-truth generation and the \textit{model-side causal path} for prediction leads to stronger performance and more effective personalization.


\begin{table}[t]
\centering
\caption{Ablation results on personalized text generation across three benchmark datasets. ``RI'' ($\%$) refers to the relative improvement of each ablation variant over the Base Model.}
\resizebox{\textwidth}{!}{
\fontsize{18}{30}\selectfont
\begin{tabular}{ccccccccccccc}
\toprule
\textbf{Datasets ($\rightarrow$)}     & \multicolumn{4}{c}{{\textbf{Book Review}}}& \multicolumn{4}{c}{{\textbf{Movie Review}}}& \multicolumn{4}{c}{{\textbf{CD Review}}}\\ \cmidrule(lr){2-5} \cmidrule(lr){6-9} \cmidrule(lr){10-13}
\textbf{Methods ($\downarrow$)}     & ROUGE-1     &RI      & ROUGE-L & RI & ROUGE-1     &RI      & ROUGE-L & RI  & ROUGE-1     &RI      & ROUGE-L & RI \\ \midrule
\textbf{Base Model (SFT)}   & {0.0752}    & {-}    & {0.0351} & {-}    & {0.0620}    & {-}  & {0.0305}    & {-}    & {0.0668}  & {-}    & {0.0314}    & {-} \\

\textbf{$\textbf{+}$ MCEOnly}& 0.1827  &  {142.9 \textcolor{red}{$\uparrow$}} & 0.0907 & {158.4 \textcolor{red}{$\uparrow$}}  & 0.1629  & {162.7 \textcolor{red}{$\uparrow$}} & 0.0796 & {161.0 \textcolor{red}{$\uparrow$}}  & 0.1552 & {132.3 \textcolor{red}{$\uparrow$}}  & 0.0744  &  {136.9 \textcolor{red}{$\uparrow$}}\\

{\textbf{$\textbf{+}$ MCE-DCE Alignment}} & {0.1876}  &  {{149.5} \textcolor{red}{$\uparrow$}} & {0.0961} & {{173.6} \textcolor{red}{$\uparrow$}}  & {0.1671}  & {{169.5} \textcolor{red}{$\uparrow$}} & {0.0817} & {{168.0} \textcolor{red}{$\uparrow$}}  & {0.1672} & {{150.3} \textcolor{red}{$\uparrow$}}  & {0.0793}  &  {{152.5} \textcolor{red}{$\uparrow$}}\\

\textbf{$\textbf{+}$ DCE\_Only}     & {0.1958}    & {160.4 \textcolor{red}{$\uparrow$}}    & 0.1122   & {219.7 \textcolor{red}{$\uparrow$}}    & {0.1865}    & {200.8 \textcolor{red}{$\uparrow$}} & {0.0953} & {212.4 \textcolor{red}{$\uparrow$}}    & {0.1805} & {170.2 \textcolor{red}{$\uparrow$}}    & {0.0922} & {193.6 \textcolor{red}{$\uparrow$}}   \\
\textbf{$\textbf{+}$ Full (\methodnameshort)}   & {{\textbf{0.2318}}}    & {208.2  \textcolor{red}{$\uparrow$}}   & {{\textbf{0.1270}}} &  {261.8 \textcolor{red}{$\uparrow$}}  & {{\textbf{0.2015}}}    & {225.0 \textcolor{red}{$\uparrow$}}   & \textbf{0.1041}   & {241.3 \textcolor{red}{$\uparrow$}} & \textbf{0.1976}    & {195.8 \textcolor{red}{$\uparrow$}} & {\textbf{0.0998}}   & {217.8 \textcolor{red}{$\uparrow$}} \\ \bottomrule   
\end{tabular}
}
\label{tab: ablation}
\end{table}

\subsection{In-depth Analysis}
\paragraph{Ablation Studies. }
To better understand the contribution of each component in our framework, we conduct an ablation study focusing on two key design choices: \textit{causal preference modeling} and \textit{token-level preference weighting}.  
We construct several ablation variants by selectively disabling these components (see Appendix~\ref{appendix:ablation-variants} for implementation details). 

{As shown in Table ~\ref{tab: ablation}, incorporating DCE (\textsc{DCEOnly}), MCE
(\textsc{MCEOnly}), or MCE-DCE alignment (\textsc{MCE-DCE Alignment}) results in significant improvements over SFT
baseline, demonstrating that each causal component independently enhances
personalization. 
}
The full model (\methodnameshort), which integrates both components, achieves the best overall performance.
These findings validate the effectiveness of our design: modeling both \textit{what} the model learns from user history (via causal preference) and \textit{where} to focus supervision (via token weighting) is crucial for enhancing personalization.

\begin{figure}[t]
    \centering
    \begin{minipage}{0.24\textwidth}
        \centering
        \includegraphics[width=\linewidth]{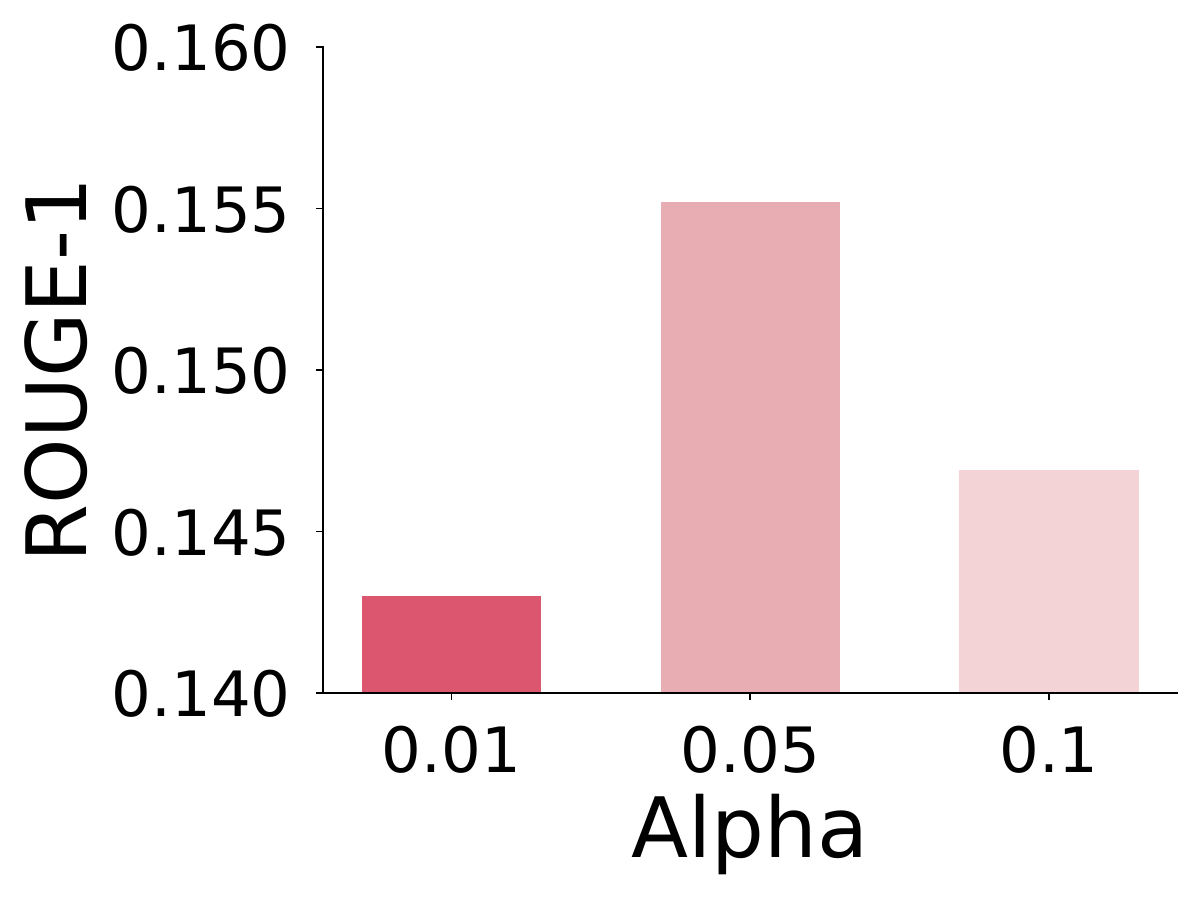}
    \end{minipage}
    \begin{minipage}{0.24\textwidth}
        \centering
        \includegraphics[width=\linewidth]{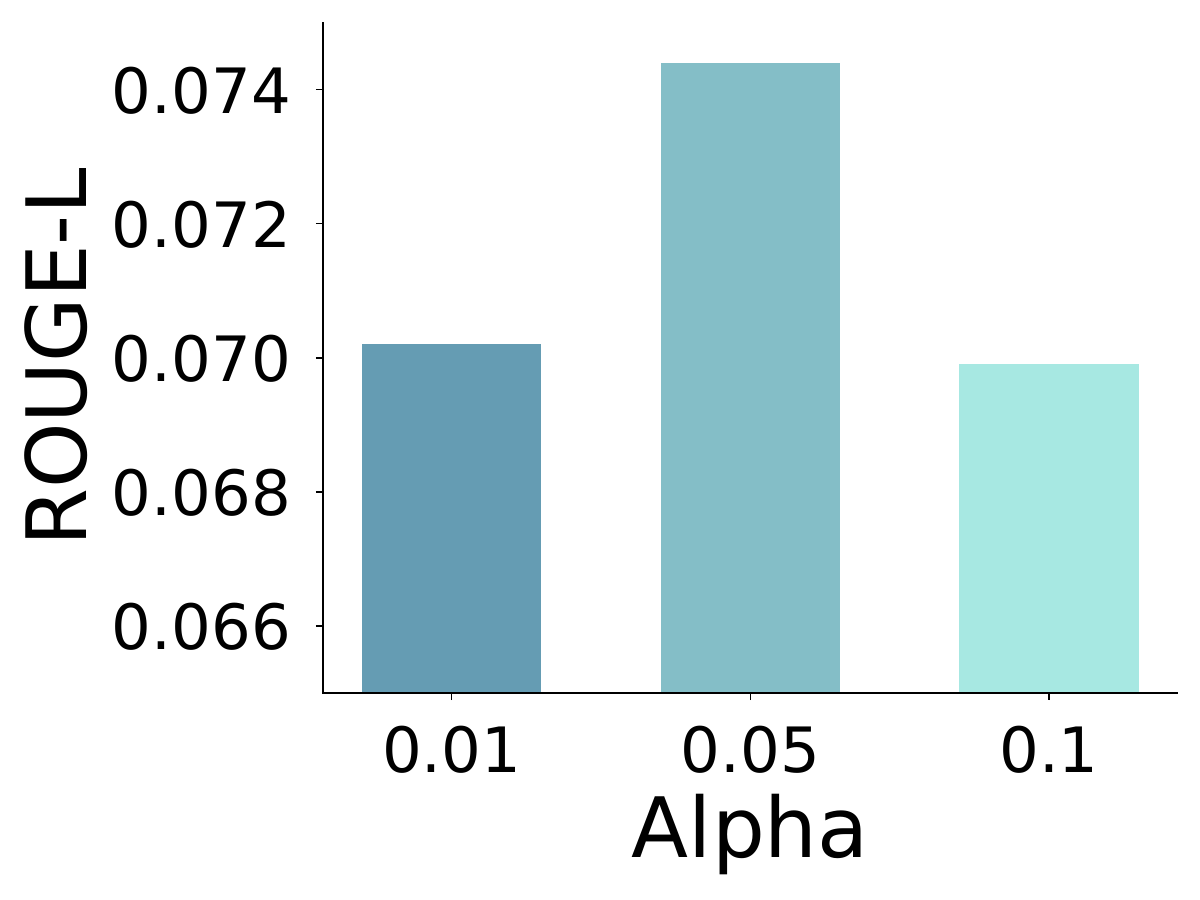}
    \end{minipage}
    \begin{minipage}{0.24\textwidth}
        \centering
        \includegraphics[width=\linewidth]{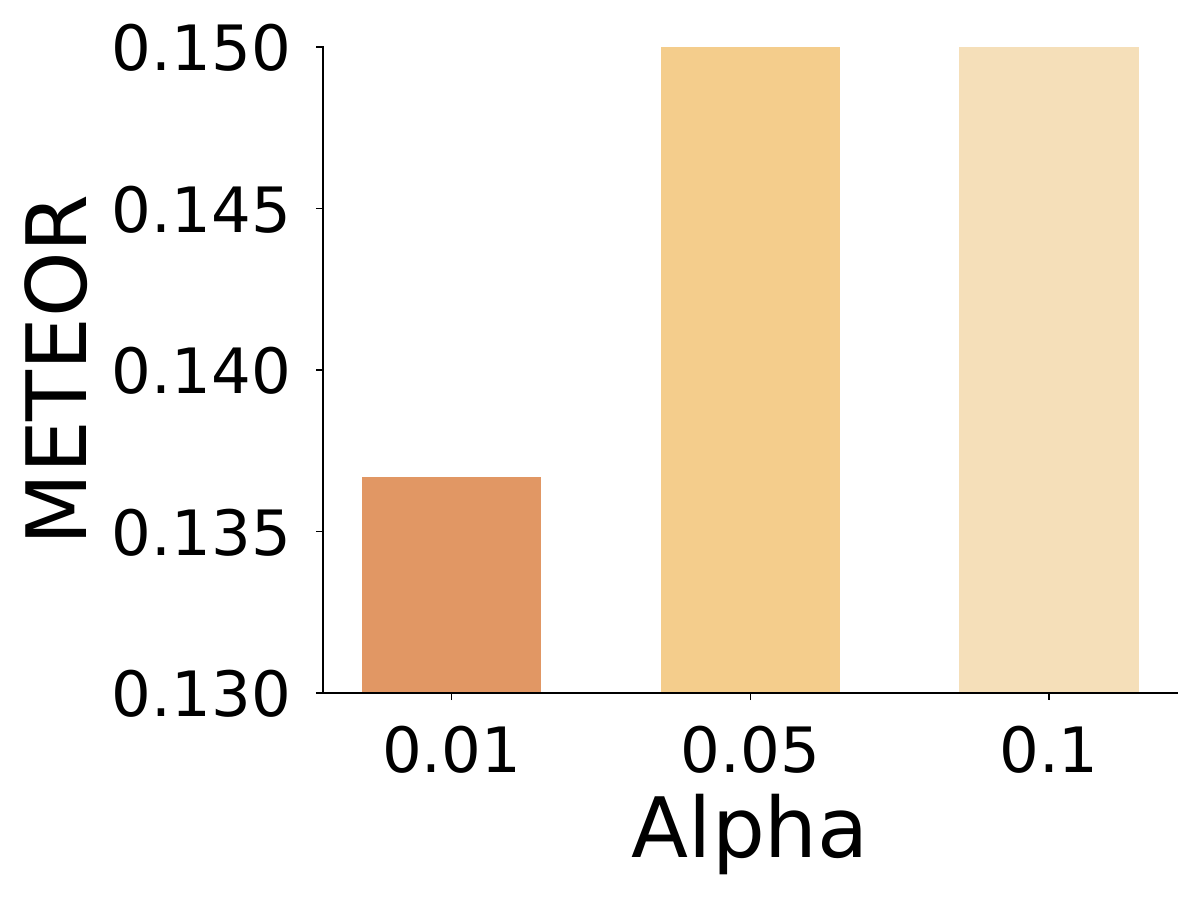}
    \end{minipage}
    \begin{minipage}{0.24\textwidth}
        \centering
        \includegraphics[width=\linewidth]{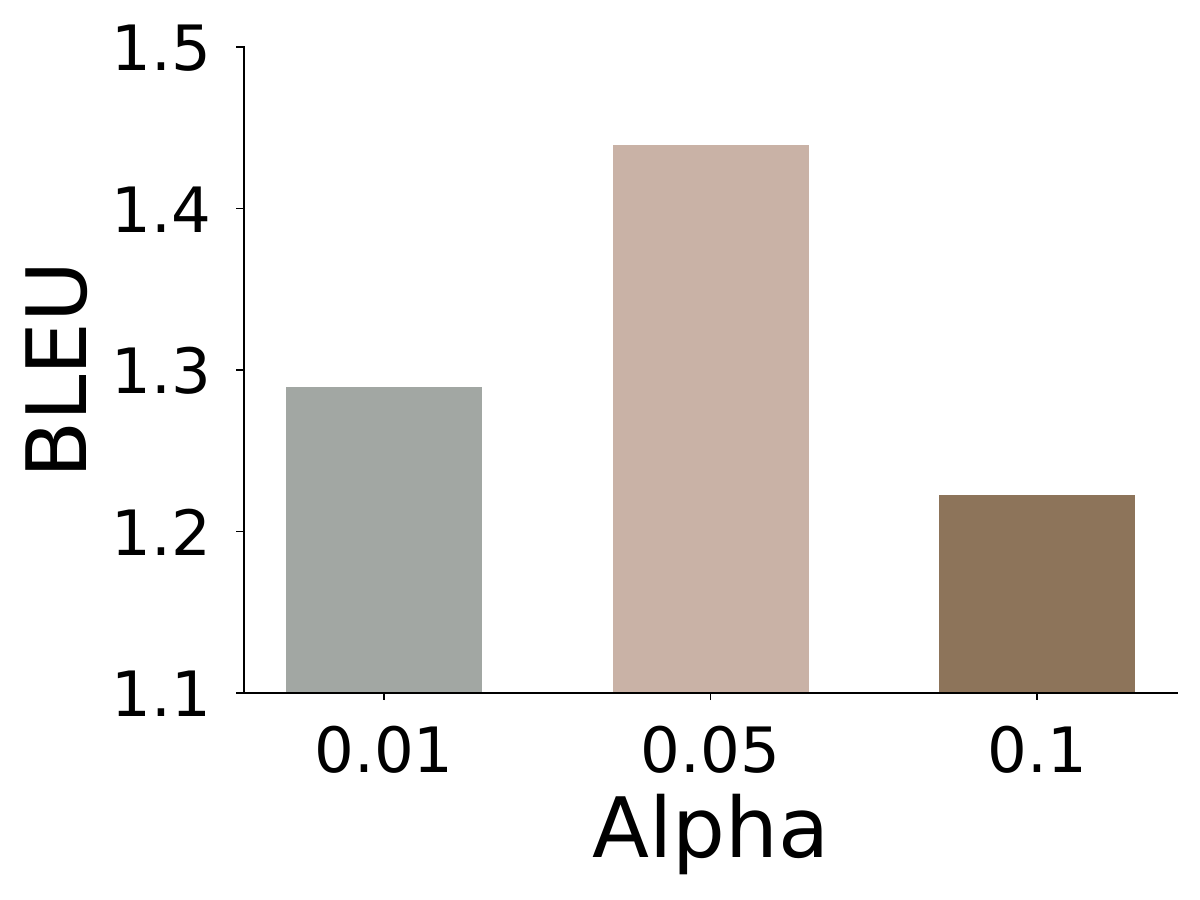}
    \end{minipage}
    
    \caption{Effect of hyper-parameter $\alpha$ on the performance.}
    \label{fig:hyper-alpha}
\end{figure}
\begin{table*}[t]
\centering
\caption{Comparison of ContextSFT and NextQuill across model sizes (1.5B, 3B, 7B), highlighting the robustness of NextQuill to model scale.}
\setlength{\tabcolsep}{15pt} 
\resizebox{\textwidth}{!}{
\begin{tabular}{clcccccc}
\toprule
\multirow{2}{*}{\textbf{Datasets}} & \multirow{2}{*}{\textbf{Metrics~($\downarrow$)}} 
    & \multicolumn{2}{c}{\textbf{1.5B}} 
    & \multicolumn{2}{c}{\textbf{3B}} 
    & \multicolumn{2}{c}{\textbf{7B}} \\ 
\cmidrule(lr){3-4} \cmidrule(lr){5-6} \cmidrule(lr){7-8}
& & \textbf{ContextSFT} & \textbf{NextQuill} 
  & \textbf{ContextSFT} & \textbf{NextQuill} 
  & \textbf{ContextSFT} & \textbf{NextQuill} \\
\midrule
\multirow{4}{*}{\parbox[c]{2cm}{\centering \textbf{Book\\Review}}}
& \textbf{ROUGE-1} & 0.1434 & 0.1631 & 0.1661 & 0.2318 & \uline{0.3879} & \textbf{0.4116} \\
& \textbf{ROUGE-L} & 0.0728 & 0.0938 & 0.0836 & 0.1270 & \uline{0.2183} & \textbf{0.2458} \\
& \textbf{METEOR}  & 0.1734 & 0.2282 & 0.2158 & 0.2580 & \uline{0.3317} & \textbf{0.3737} \\
& \textbf{BLEU}    & 2.4144 & 2.7469 & 2.1214 & 3.5718 & \uline{14.883} & \textbf{18.158} \\
\midrule
\multirow{4}{*}{\parbox[c]{2cm}{\centering \textbf{Movie\\Review}}}
& \textbf{ROUGE-1} & 0.1151 & 0.1251 & 0.1573 & 0.2015 & \uline{0.3143} & \textbf{0.3246} \\
& \textbf{ROUGE-L} & 0.0547 & 0.0637 & 0.0753 & 0.1041 & \uline{0.1526} & \textbf{0.1642} \\
& \textbf{METEOR}  & 0.1286 & 0.1512 & 0.1718 & 0.1918 & \uline{0.2477} & \textbf{0.2602} \\
& \textbf{BLEU}    & 1.5693 & 1.6808 & 1.7151 & 2.3845 & \uline{7.8608} & \textbf{9.1527} \\
\midrule
\multirow{4}{*}{\parbox[c]{2cm}{\centering \textbf{CD\\Review}}}
& \textbf{ROUGE-1} & 0.1171 & 0.1326 & 0.1505 & 0.1976 & \uline{0.3189} & \textbf{0.3238} \\
& \textbf{ROUGE-L} & 0.0561 & 0.0653 & 0.0714 & 0.0998 & \uline{0.1510} & \textbf{0.1576} \\
& \textbf{METEOR}  & 0.1173 & 0.1547 & 0.1583 & 0.1805 & \uline{0.2400} & \textbf{0.2542} \\
& \textbf{BLEU}    & 1.1502 & 1.4181 & 1.3487 & 1.9676 & \uline{7.2476} & \textbf{9.1314} \\
\bottomrule
\end{tabular}
}
\setlength{\tabcolsep}{6pt} 
\label{tab:contextsft_nextquill_results}
\end{table*}

\paragraph{Scaling Analysis.}
Table~\ref{tab:contextsft_nextquill_results} compares NextQuill with the strongest baseline ContextSFT across different backbone sizes (Qwen2.5-1.5B, 3B, 7B). We observe that NextQuill consistently outperforms ContextSFT on all datasets and metrics, with the performance gap widening as the model size increases. These results demonstrate that our causal preference alignment mechanism scales favorably with larger models, effectively leveraging additional capacity for improved personalization.

\paragraph{Hyper-parameter Analysis. }
We conduct experiments on the \textit{CD Review} dataset to analyze the impact of two key hyper-parameters in our framework: (1) $\alpha$ in Equation~\ref{eq:final-loss}, which controls the weight of the causal preference loss, and (2) $\lambda$ in Equation~\ref{eq:definition-w}, which modulates the strength of the token-level preference weighting.
To isolate their individual effects, we disable the token weighting strategy when tuning $\alpha$, and remove the weighted causal preference loss when tuning $\lambda$.
The results for $\alpha$, evaluated across multiple metrics, are shown in Figure~\ref{fig:hyper-alpha}. We find that $\alpha=0.05$ consistently yields strong performance and serves as a robust setting across datasets. The detailed analysis of $\lambda$ is reported in Appendix~\ref{appendix:hyper-param}.



\begin{figure}[t]
    \centering
    \subfigure[Book Review]{\includegraphics[width=0.31\textwidth]{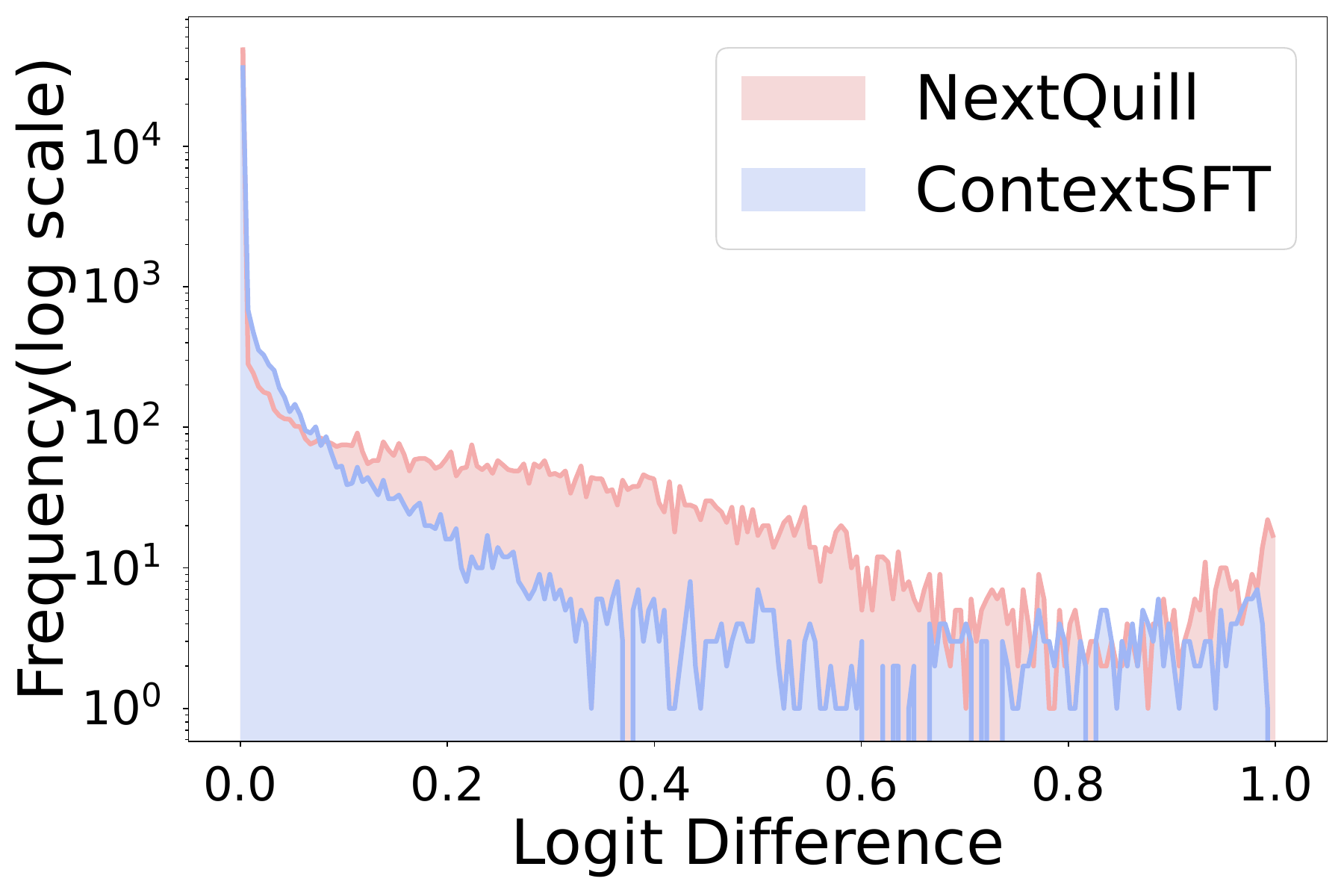}}
    \subfigure[Movie Review]{\includegraphics[width=0.31\textwidth]{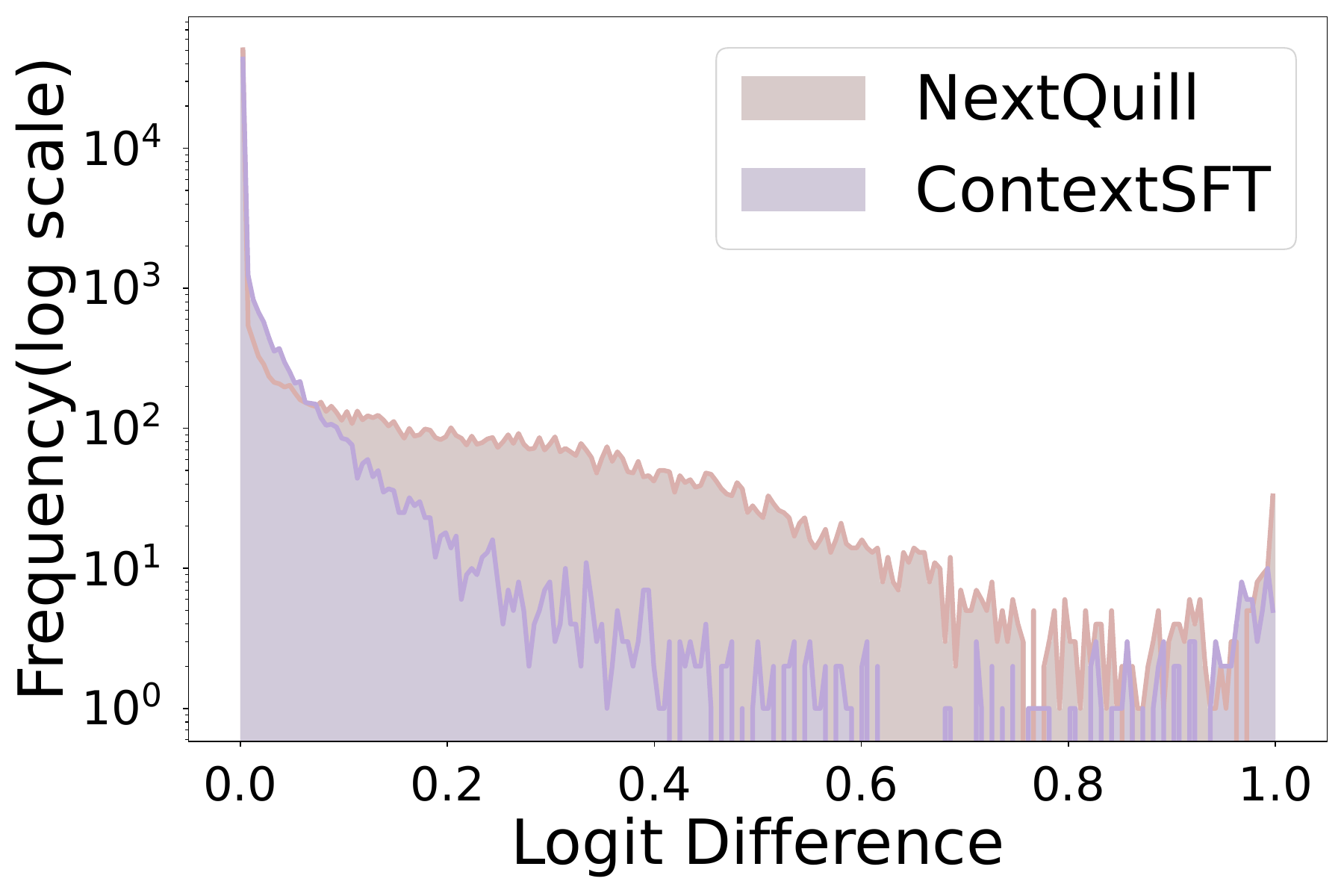}}
    \subfigure[CD Review]{\includegraphics[width=0.31\textwidth]{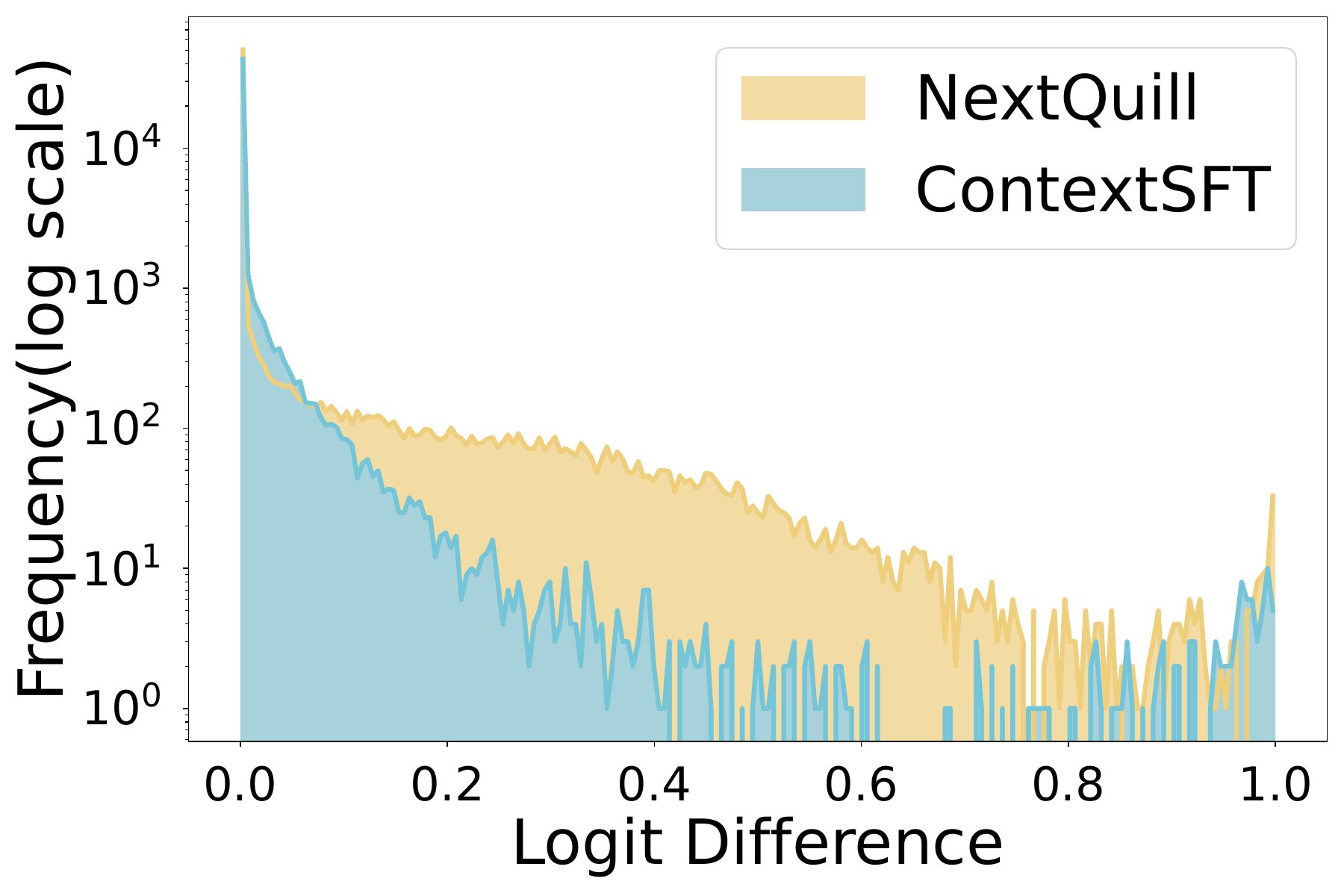}}
    \vspace{-10pt}
    \caption{Logit difference distributions with and without user history across three benchmark datasets.}
    \label{fig:Diff}
\end{figure}
\subsection{Word-Level Preference Analysis. }
To further validate the ability of \methodnameshort to capture user preferences, we conduct an quantitative analyse at the token level. Specifically, we randomly sample 50 training instances from each dataset and compare the difference in token-level logits with and without user history for both \methodnameshort and ContextSFT. As shown in Figure~\ref{fig:Diff}, \methodnameshort produces significantly larger logit differences, indicating that it learns stronger preference representations. This confirms that our model captures more user-specific signals, which in turn helps better align the generated output with personalized aspects of the target text. 

\section{Related Work}

\paragraph{LLM Personalization.} 
 {LLM personalization has gained increasing attention across text~\citep{zhao2025exploring,zhang2025personalized,PDRec}, conversation~\citep{li2025personalized,RSO} and multimodal generation~\citep{pmg}. Recent Benchmarks such as LaMP~\citep{salemi2024lamp} and LongLaMP~\citep{kumar2024longlamp} provide standardized settings for training and evaluating personalized text generation.
Existing methods mainly follow two paradigms: retrieval-based prompting and fine-tuning with user history.
Retrieval-based approaches~\citep{ropg,hydra} augment model inputs by retrieving user-specific context from external memory. DPL~\citep{DPL} identifies inter-user differences to improve personalization. SteerX~\citep{steerX} leverages causal inference to disentangle preference-driven from preference-agnostic tokens and uses these signals to steer generation.
Fine-tuning-based approaches adapt model parameters using user histories. Parameter-efficient methods~\citep{perpcs,personpeft} achieve scalable personalization across users. ONCE~\citep{once} constructs profile summaries from behavioral logs, and PPlug~\citep{pplug} encodes user histories into personalized embeddings. P-RLHF~\citep{li2024personalized} jointly learns a user model and a personalized LLM from human feedback. DEP~\citep{DEP} further models inter-user differences in latent space rather than via prompt signals.
However, most existing methods treat all tokens uniformly, lacking mechanisms to identify which parts of the input or output truly reflect user intent. In contrast, we introduce a causal preference modeling framework that attributes token-level effects through counterfactual reasoning, enabling principled supervision over both model-internal representations and output behavior to achieve genuine personalization.}

\paragraph{Causal Inference for LLMs.}
Causal inference provides a principled framework for modeling cause-effect relationships in both observational and interventional settings~\citep{peters2017elements}. With the rise of LLMs, causal methods have been increasingly adopted to better understand, analyze, and improve model behavior across a range of NLP tasks \citep{stolfo2023causal,zhu2024mitigating,yu2024cause}.
A growing body of research investigates whether LLMs possess causal reasoning abilities \citep{jin2024can}, and explores how causal inference can be applied when treating text as either the treatment or the outcome variable \citep{zhou2024causalbench,veitch2020adapting}. Other work leverages causal reasoning to improve LLM capabilities in domains such as recommendation systems \citep{zhang2021causal,zhang2023leveraging} and arithmetic reasoning \citep{chi2024unveiling,berg2020empirical,tseng2024two}. 
Despite these advances, causal techniques have not yet been systematically applied to the challenge of LLM personalization. To our knowledge, this work is the first to propose a unified causal framework that models user preference effects from both the model side and the data side, enabling fine-grained attribution and alignment for personalized text generation.

\section{Conclusion}
We introduce \methodnameshort, a novel causal preference modeling framework for LLM personalization that attributes and aligns user-specific signals through token-level causal effect estimation. By explicitly modeling both model-side and data-side preference pathways, \methodnameshort enables targeted, interpretable supervision aligned with true user preferences.
Unlike prior methods that uniformly fit all predictions and target tokens, our approach aligns model-internal causal effects with preference-bearing supervision signals, allowing the model to learn \textit{what} to adapt and \textit{where} to focus.
Experiments across diverse benchmarks confirm the effectiveness of 
\methodnameshort in improving personalized generation performance. We provide a detailed discussion of the limitations in Appendix~\ref{appx:limitation}.

\section*{Ethics Statement}
This research adheres to the ICLR Code of Ethics. It does not involve human subjects or sensitive personal information. 
We use only user data that individuals are willing to publicly disclose, \textit{e.g.}, reviews or posts. Moreover, all user information is anonymized, containing no personally identifiable data, and our use of the dataset strictly adheres to the guidelines established by the data providers. 
Although our experiments and method design do not pose direct risks of harm, we acknowledge that applying our method in private domains should involve stronger privacy safeguards, such as providing users with control over their data and supporting local learning scenarios.
Details regarding the use of large language models (LLMs) are provided in Appendix~\ref{appendix:LLMuse}.

\section*{Reproducibility statement}
To ensure reproducibility of our work, we provide detailed descriptions of our method, model architectures, and training procedures in Sections~\ref{sec:method} and~\ref{experiment_label} of the main text. 
All hyperparameters, data preprocessing steps, and evaluation protocols are documented in Section~\ref{sec:expsetting} and Appendix~\ref{appendix:implementation}. Additionally, we include pseudo-code for our algorithm in Appendix~\ref{appendix:alg} and provide links to our code and datasets (\url{https://github.com/juntaoyou/NextQuill}). 

\subsubsection*{Acknowledgments}
This work is supported by grants from the Research Grants Council of the Hong Kong Special Administrative Region, China (No. CUHK 14206625).

\bibliography{iclr2026_conference}
\bibliographystyle{iclr2026_conference}

\appendix
\input{appendix}

\end{document}

%% file: math_commands.tex
\usepackage{amsmath,amsfonts,bm}

\def\eqref#1{equation~\ref{#1}}

\def\1{\bm{1}}

\DeclareMathAlphabet{\mathsfit}{\encodingdefault}{\sfdefault}{m}{sl}
\SetMathAlphabet{\mathsfit}{bold}{\encodingdefault}{\sfdefault}{bx}{n}

\def\sD{{\mathbb{D}}}



%% file: 1_intro.tex
\section{Introduction}
\label{intro}
Large Language Models (LLMs) have exhibited exceptional capabilities across various domains ~\citep{gpt4,zhao2024comprehensive}, driving their widespread deployment in real-world applications ~\citep{userinterestjourney,liu2025inference,shi2024large,zhao2026don} such as virtual assistants and content generation systems. However, existing LLMs are typically designed under a general-purpose ``one-size-fits-all'' paradigm~\citep{DPL}, neglecting the diversity of user preferences in aspects such as needs, goals, and communication styles. 
As LLMs become more integrated into individuals’ daily lives and work, accounting for personal preferences in LLMs is becoming increasingly important for delivering tailored and engaging experiences~\citep{pearl, liu2025survey}. This growing demand has sparked a surge of research interest in \textit{LLM personalization}, with notable efforts emerging across both academic research and industry~\citep{myvlm, plvm, teachllm}.

Intuitively, user historical data inherently reflects individual preferences, playing a critical role in enabling personalization for LLMs. 
Based on how this data is utilized, existing personalization approaches can be broadly classified into two categories. The first follows a memory-retrieval paradigm~\citep{ropg,DPL,hydra}, in which user history is stored in an external memory and relevant information is then dynamically retrieved and injected into the prompt to guide generation. 
In this setting, LLMs rely on steering prompts to produce user-aligned responses.
However, performance may be limited due to the lack of alignment mechanisms specifically designed for personalization. 
The second category involves further fine-tuning LLMs with historical data to adapt model parameters~\citep{instructrec, personpeft, pplug} --- typically, tuning the model to leverage past user behaviors for predicting subsequent ones.
This fine-tuning process customizes the model's capacity to capture individual preferences, facilitating more explicit personalization alignment.

Although existing tuning-based methods improve alignment, we argue that their alignment remains suboptimal due to superficial modeling and the uncritical use of user data, overlooking what truly matters for preference modeling. 
From the perspective of preference representation within the model, these methods typically treat all predictions generated from the entire input as inferred preferences and align them uniformly with the ground truth. This overlooks the fact that it is primarily the inferences derived from historical behavioral data that genuinely reflect the model’s internal preference modeling. 
From the data supervision perspective, these methods treat all tokens in the ground-truth responses equally, failing to account for the unequal contribution of different tokens to the expression of user preferences. Such shallow treatment prevents the model from identifying and emphasizing the preference-driven components critical for personalized generation, limiting the alignment quality.

\begin{figure}[t]
    \centering
    \includegraphics[width=\textwidth]{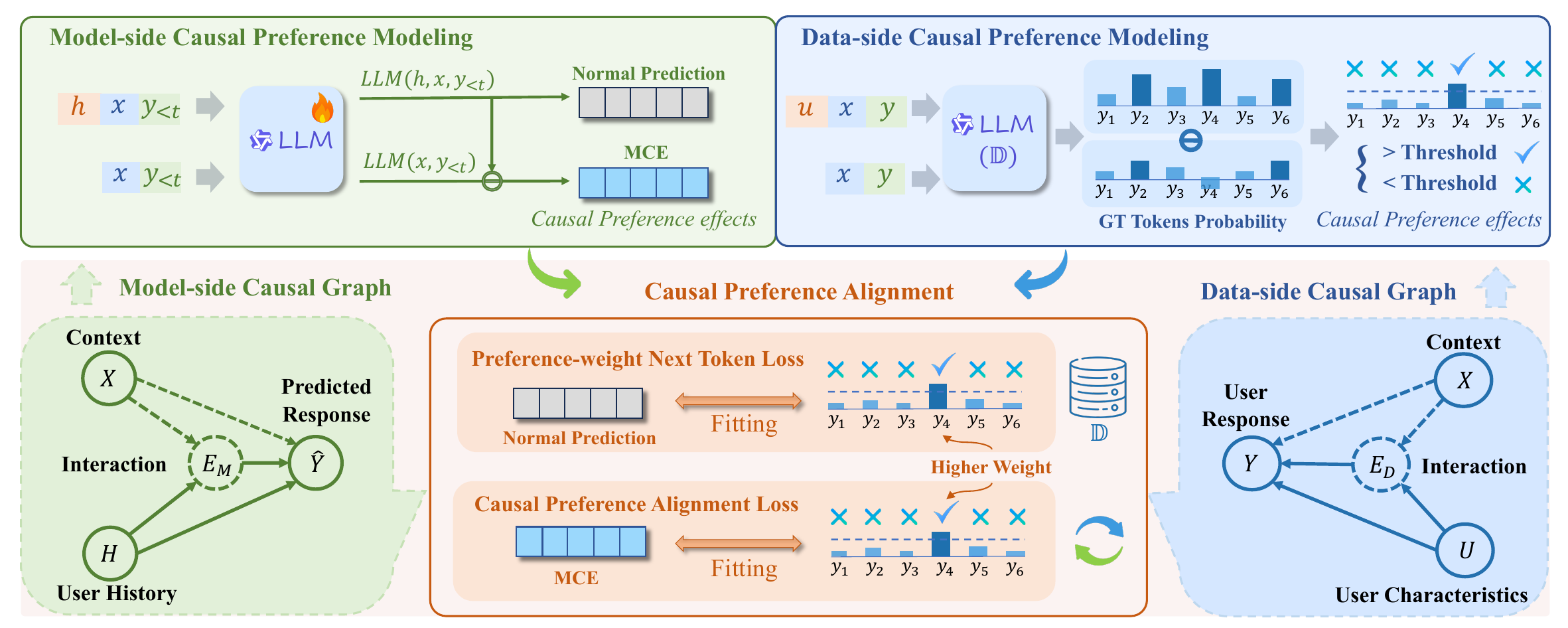}
   \caption{Illustration of NextQuill. The model-side causal graph depicts the response prediction process of LLMs, in which the outcome is influenced by user history, context information, and their interaction. The data-side causal graph describes how users generate the ground-truth response, which is jointly influenced by user characteristics, the context, and their interaction. 
   Based on these, NextQuill introduces the causal preference modeling, defining causal preference effects on both model and data sides, and accordingly introduces two causal preference alignment strategies (\textit{i.e.,} two losses) to enhance the personalization of LLMs.
    }
    \label{fig:intro}
\end{figure}


To address the issues, we must determine what truly matters in preference modeling. We approach this from a causal perspective, treating the generation of predicted responses (by model) and the creation of ground-truth responses (by users) as outcomes of an underlying causal process, separately depicted by the left and right causal graphs in Figure~\ref{fig:intro}. 
As shown, both the generation of model predictions and the creation of ground-truth responses can be influenced by non-preference factors—beyond user history for model predictions and beyond user characteristics for ground-truth responses, mainly contextual factors.
However, in theory, only the component caused by preference-related factors can accurately represent the user preference. 
To isolate the preference-driven components on both the model and data sides, we propose Causal Preference Modeling, grounded in causal effect estimation. Specifically, on the model side, we define a causal preference effect as the causal effect of historical user data on model predictions,
representing the true internal preferences encoded by the LLM. On the data side, we define a causal preference effect as the causal effect of user characteristics on the creation of ground-truth responses,
measuring the extent to which each token in the ground-truth is driven by user preferences.


Building on this foundation, we propose \textit{\methodnameshort}, a novel causal preference modeling-based alignment method for LLM personalization. \methodnameshort introduces two strategies to highlight preference-driven components during alignment on both the model and data sides.
On the model side, we use causal preference effects to isolate the preference-driven components of model predictions, representing the preferences internalized by the LLM, and then introduce a new loss function to align these effects with those in the ground truth—rather than indiscriminately aligning all predictions.
On the data side, we utilize causal preference effects to define a causal attribution score for the ground-truth data tokens, identifying those that are truly driven by user preferences. 
We then assign higher attention weights to these tokens during data fitting, ensuring that preference alignment focuses on the most relevant, preference-driven tokens. 
By integrating these two strategies, \methodnameshort shifts the alignment process toward learning from causal preference effects, enabling more effective and personalized LLM adaptation.

The main contribution of this work can be summarized as follows:
\begin{itemize}[leftmargin=*]
    \item 
    We introduce a causal perspective on LLM personalization and propose a causal preference modeling approach, defining causal preference effects at both model and data sides.
    \item We propose \methodnameshort, a causal preference modeling-based alignment method that enhances LLM personalization by 1) aligning model-internal causal preference effects with those in ground-truth response,
    rather than indiscriminately aligning all predictions 
    and 2) focusing learning on preference-driven ground-truth tokens, 
    rather than treating all tokens uniformly.
    \item We conduct comprehensive experiments across multiple personalization domains, showing that \methodnameshort significantly improves generation quality and overall personalization performance.
    
\end{itemize}


%% file: 3_method.tex
\section{Methodology}
\label{sec:method}

To improve LLM personalizaiton, we must determine what truly matters in preference modeling. At the beginning of the section, we conduct a causal analysis to address this question, introducing a new perspective of causal preference modeling. Following this, we present our \methodnameshort method, which explicitly emphasizes causal preference effects on both the model and data side during training to improve LLM personalization.


\subsection{Causal Preference Modeling}
\label{sec:causalpreference}
The core of causal preference modeling lies in analyzing the processes of response prediction by the model (model-side) and ground-truth response generation by users (data-side) using \textit{causal graph}~\citep{pearl2009causality}, deriving the causal effects related to preference factors as the basis for preference modeling. We next provide the analysis from the two sides, respectively.

\subsubsection{Model-side Causal Preference Modeling}
By definition, a \textit{causal graph} is a directed acyclic graph (DAG), where each node represents a variable and each edge indicates a causal relationship between two variables~\citep{zhou2024causalbench}. It serves as a powerful tool for describing the processes of data generation or model prediction, thereby guiding method design~\citep{zhang2021causal}.


\noindent \textbf{Model-side Causal Graph.} In our work, we use the left causal graph in Figure~\ref{fig:intro} to illustrate the response prediction process of LLMs, comprising the following nodes:
\begin{itemize}[left=0pt, itemsep=0pt]
\item $H$: Historical user data that reflects user preferences. 

\item $X$: Other context information, \textit{e.g.,} query and task prompt.

\item $E_{M}$: The latent variable that captures the interaction between $X$ and $H$ in the model.

\item $\hat{Y}$: The predicted response.
\end{itemize}
Due to the complex prediction pathway in the model, $H$ and $X$ can each independently influence the predicted response or interact with each other to do so, as represented by the following edges:
\begin{itemize}[left=0pt, itemsep=0pt]
\item $H \rightarrow \hat{Y}$ \& $X \rightarrow \hat{Y}$: $H$ and $X$ can independently affect $\hat{Y}$.
\item $(H, X) \rightarrow E_{M} \rightarrow \hat{Y}$: $H$ and $X$ interact with each other to affect $\hat{Y}$.
\end{itemize}
As shown, both $H$ and $X$ exert causal effects on $\hat{Y}$. However, it is the effects of $H$—driven by user information—that are most indicative of preference signals. Therefore, we need to identify these causal effects as the key to effective preference modeling, which we term model-side causal preference effects. 

\paragraph{Model-side Causal Preference Effect.}
Based on causal theory and the corresponding causal graph, the target causal effect of $H=h$ (conditioned on a given $x$) can be formulated as:
\begin{equation}\label{eq:CE}\small
\begin{split}
    \textit{MCE}(\hat{Y}_t|h,x) & = P(\hat{Y}_t \mid H = do(h), x) - P(\hat{Y}_t \mid H = do(0), x) \\
       & = P(\hat{Y}_t \mid H = h, x) - P(\hat{Y}_t \mid H = 0, x),
\end{split}
\end{equation}
where $MCE(\hat{Y}_t|h,x)$ denotes the model-side casual effects, $\hat{Y}_t$ denotes the $t$-th token for the prediction, and $do(\cdot)$ denotes the $do$-calculus. The expression $H = do(h)$ represents an intervention that sets $H$ to the specific value $h$, while $H = do(0)$ denotes setting $H$ to a reference value (\textit{e.g.}, a null). The term $P(\hat{Y}_t \mid H = do(h), x)$ refers to the interventional probability, whereas $P(\hat{Y}_t \mid H = h, x)$ is the corresponding observational (Bayesian) probability. Under our causal graph, these two are equivalent, \textit{i.e.,} $P(\hat{Y}_t \mid H = do(h), x) = P(\hat{Y}_t \mid H = h, x)$, and similarly for the reference case.

\textbf{Interpretation:} Functionally, this model-side causal effect measures the part of the predictions that is truly driven by the preferences represented by the user’s history, \textit{reflecting the true internal preferences modeled (or captured) by LLMs} from inputs. This suggests that not all parts of the model's predictions should be treated equally.

\subsubsection{Data-side Causal Preference Modeling}

\noindent \textbf{Data-side Causal Graph.} The right causal graph in Figure~\ref{fig:intro} depicts the ground-truth response generation process (\textit{i.e.}, how users produce the ground-truth response). The rationale for this causal graph is explained as follows:

\begin{itemize}[left=0pt, itemsep=0pt]
\item $U$: User characteristics, which indicate the user’s underlying preference characteristics.

\item $X$: Context information, mainly the item or the objective the user focuses on.

\item $E_{D}$: The latent variable that captures the interaction between $U$ and $X$.

\item $Y$: The outcome variable, representing the user-written ground-truth response.
\end{itemize}

The causal relationships between these variables are described by the following edges:
\begin{itemize}[left=0pt, itemsep=0pt]

\item $(U, X) \rightarrow E_{D} \rightarrow Y$: $U$ and $X$ can influence $Y$ through their interaction $E_{D}$, indicating that when generating $Y$, the user engages with the context $X$ to focus on the parts of interest.

\item $X \rightarrow Y$: $X$ can influence $Y$ independently of $U$; for instance, no matter who the user is, the lead actor is often mentioned in the reviews for a specific movie.

\item $U\rightarrow Y$: $U$ can influence $Y$ independently of $X$. For example, regardless of the content $X$, a user may consistently use the same catchphrases. 
\end{itemize}

Similarly, both $U$ and $X$ have causal effects on $Y$, but it is the effects of $U$—reflecting user preferences—that most directly convey preference signals. Identifying $U$’s causal influence on $Y$ is therefore essential for isolating preference in the ground-truth response. We refer to these causal effects as the data-side causal preference effects.

\paragraph{Data-side Causal Preference Effect.}  

Similar to the model side, the data-side causal effect of $U=u$ on $Y$ (conditioned on a given $x$) can be formulated as:
\begin{equation}\label{eq:DCE}\small
\begin{split}
    DCE(Y_t|u,x) & = P(Y_t \mid U = do(u), x) - P(Y_t \mid U = do(0), x) \\
       & = P(Y_t \mid U = u, x) - P(Y_t \mid U = 0, x),
\end{split}
\end{equation}
where $Y_{t}$ denotes the $t$-th token of $Y$, $DCE(Y_t \mid u, x)$ denotes the obtained data-side causal effects, and $u$ represents the characteristics of the specific user. 

\textbf{Analysis}: Given a sample $(x,h,y) \in \sD$, by specifying $Y_{t}$ as the $t$-th token of $y$ (\textit{i.e.}, $y_{t}$), the causal effect $DCE(Y_t = y_t \mid u, x)$ \textit{quantifies the extent to which the generation of $y_t$ is driven by user preferences}. A higher value indicates that the token more strongly reflects the user’s preferences. This implies that different tokens vary in their preference relevance and should be treated differently during training.


\subsection{Personalization Alignment based on Causal Preference Modeling}
Building on our analysis of the causal preference modeling framework, we find that different parts of both model predictions on the response and ground-truth response vary in their relevance to user preferences. To enhance personalization alignment, it is crucial to identify the preference-driven components based on causal preference effects and explicitly emphasize their learning. Specifically, once identified, we adopt two key strategies to strengthen their influence: 1) on the one hand, we prioritize learning from ground-truth tokens that are driven by user preferences; 2) on the other hand, we introduce a causal preference alignment loss to enforce alignment between the true internal preferences modeled by LLMs (model-side causal preference effects) and those in the ground-truth data (data-side causal preference effects). 
Figure~\ref{fig:intro} provides an overview of our method.
In the following, we first describe how to identify preference-driven components, followed by the tuning strategies. 

\subsubsection{Preference-driven Component Identification}
Section~\ref{sec:causalpreference} provides key insights for identifying preference-driven components in both model predictions and ground-truth responses. On the model side, we can use the causal preference effects on predictions to extract the parts truly reflecting user preferences. On the data side, we can assess the causal preference effects for each ground-truth token to determine whether it is preference-driven.
\paragraph{Identification on Model Side.}
On the model side, we leverage the causal preference effect of user history on predictions, as $MCE$ defined in Equation~\ref{eq:CE}, to identify the components of predictions that reflect user preferences. The causal effect in Equation~\ref{eq:CE} is defined from a probabilistic perspective, and it must be converted into an empirical form for practical application. Given a data sample $(x, h, y) \in \sD$, the empirical causal effect for the prediction of the $t$-th token is computed as:
\begin{equation}\label{eq:prediction-CE}\small
     f_\theta(x, h, y_{<t}) - f_\theta(x, \emptyset, y_{<t}),
\end{equation}
where: 1) $f_\theta(x, h, y_{<t})$ denotes the model’s prediction for the $t$-th token, conditioned on the user query $x$, user history $h$, and the preceding tokens $y_{<t}$; and 2) $f_\theta(x, \emptyset, y_{<t})$ denotes the prediction when the user history is removed. Besides, $\theta$ denotes the learnable model parameters of LLMs.
Notably, this formulation differs slightly from Equation~\ref{eq:CE} in that we include $y_{<t}$ as part of the input. This is because $y_{<t}$ can be considered part of the evolving context (or extended query) during generation of the $t$-th token. For simplicity, we do not merge $x$ and $y_{<t}$ explicitly in notation.

\paragraph{Identification on Data Side.} 

Given a sample $(x, h, y) \in \sD$, the causal effect $DCE(Y_t = y_t \mid u, x)$ in Equation~\ref{eq:DCE} measures the extent to which the generation of the $t$-th token in $y$ (\textit{i.e.}, $y_t$) is driven by user preferences. Based on this, we classify each ground-truth token $y_t$ as either preference-driven or not. To represent this distinction, we assign a token weight $\omega_t$ to denote the results. Specifically,
\begin{equation}\label{eq:definition-w}\small
\omega_t = \begin{cases}
\lambda, & \text{if } DCE(Y_t = y_t \mid u, x) > \delta \\
\epsilon, & \text{otherwise}
\end{cases},
\end{equation}
where $\delta$ denotes the threshold. If $\omega_t = \lambda$, the token is a preference-driven token; otherwise, it is treated as a non-preference-driven token. $\lambda$ and $\epsilon$ are two manually defined weights, which will be used later in the tuning process.

\textit{Estimating $DCE(Y_t = y_t \mid u, x)$.} To estimate $DCE(Y_t = y_t \mid u, x)$ for a user with characteristics $u$, we use an LLM that has seen dataset $\sD$ for approximation in a way similar to Equation~\ref{eq:prediction-CE}. 
Moreover, since the original user characteristics $u$ are not directly available, we represent them using the historical data $h$. The estimation is then given by:
\begin{equation}\label{eq:definition-ce} \small 
DCE(Y_t = y_t \mid u, x) \approx f_{\theta_{\sD}}(y_t|x, h, y_{<t}) - f_{\theta_{\sD}}(y_{t}|x, \emptyset, y_{<t}),
\end{equation}
where $\theta_{\sD}$ denotes the model parameters of the LLM having seen $\sD$. Here, $f_{\theta_{\sD}}(y_t \mid x, h, y_{<t})$ estimates the probability that the $t$-th token is $y_t$. Since the LLM has seen $\sD$, $f_{\theta_{\sD}}(y_t \mid x, h, y_{<t})$ can serve as an approximated estimation for $P(Y_t \mid U=h, X)$ in Equation~\ref{eq:DCE}. {Appendix~\ref{app:DCE_estimation} discusses two practical implementations, including both dynamic and fixed DCE estimation.}




\subsubsection{Tuning Strategy}
To enhance learning on preference-driven components in both model predictions and ground-truth data, we adopt two strategies: (1) assign higher weights to preference-driven tokens during ground-truth response fitting, and (2) introduce a new causal preference alignment loss to align model-side causal preference effects and the data-side ones, {\textit{i.e.,} align MCE and DCE}. The weighting mechanism is applied to both the original standard next token prediction loss and the new causal preference alignment loss. We finally combine the two losses to fine-tune the model like SFT.

\paragraph{Preference-weighted Next Token Loss.} We adopt the normal next-token prediction loss to ensure coherent and appropriate text generation, while modifying it into a preference-weighted form to better emphasize the learning of preference-driven tokens, guided by the {DCE-driven} weights $\omega_t$ defined in Equation~\ref{eq:definition-w}. Specifically, the weighted loss $L_n$ is formulated as:
\begin{equation}\small
    L_n = \frac{1}{|\sD|} \sum_{(x, h, y) \in \sD} \sum_{t=1}^{|y|} \omega_t \cdot \ell(f_\theta(x, h, y_{<t}), y_t),
    \label{eq:noraml-loss}
\end{equation}
where $\ell(\cdot) $ is the standard cross-entropy loss, and $|y|$ is the length of $y$.

\paragraph{Causal Preference Alignment Loss.}
{To enhance personalization learning, we further align the data-side causal preference effects (DCE) with those reflected in the model (MCE), ensuring that the model’s internal preference representation matches the preferences expressed in the ground-truth response. }
To achieve this,
we introduce a new \textit{causal preference alignment loss}. 
Specifically, 
we leverage the preference-driven component of the prediction, as defined in Equation~\ref{eq:prediction-CE}, to align with the ground-truth data, placing greater emphasis on fitting the preference-driven tokens. The optimization objective $L_p$ is formulated as follows:
\begin{equation}\small
    L_p = \frac{1}{|\sD|} \sum_{(x, h, y) \in \sD} \sum_{t=1}^{|y|} \omega_t \cdot \ell\big(f_\theta(x, h, y_{<t}) - f_\theta(x, \emptyset, y_{<t}); y_t\big).
    \label{eq:5}
\end{equation}
This loss directly encourages alignment between the LLM's internal causal preference effects and the causal preference effects in the ground-truth.

\paragraph{Final Optimization Objective.}
We combine the two losses to fine-tune the model in a manner similar to SFT. The optimization objective is formulated as:
\begin{equation}
    \theta^{\star} = \textit{argmin}_{\theta} \quad L_n + \alpha \cdot L_p,
\label{eq:final-loss}
\end{equation}
where \( \alpha \) is a hyperparameter to balance $L_n$ and $L_p$, and $\theta^{\star}$ denotes the obtained model parameters. 


%% file: appendix.tex
\appendix
\clearpage

\section{Detailed Causal Graphs and Causal Preference Effects}

\begin{figure}[h]
    \centering
    \includegraphics[width=\textwidth]{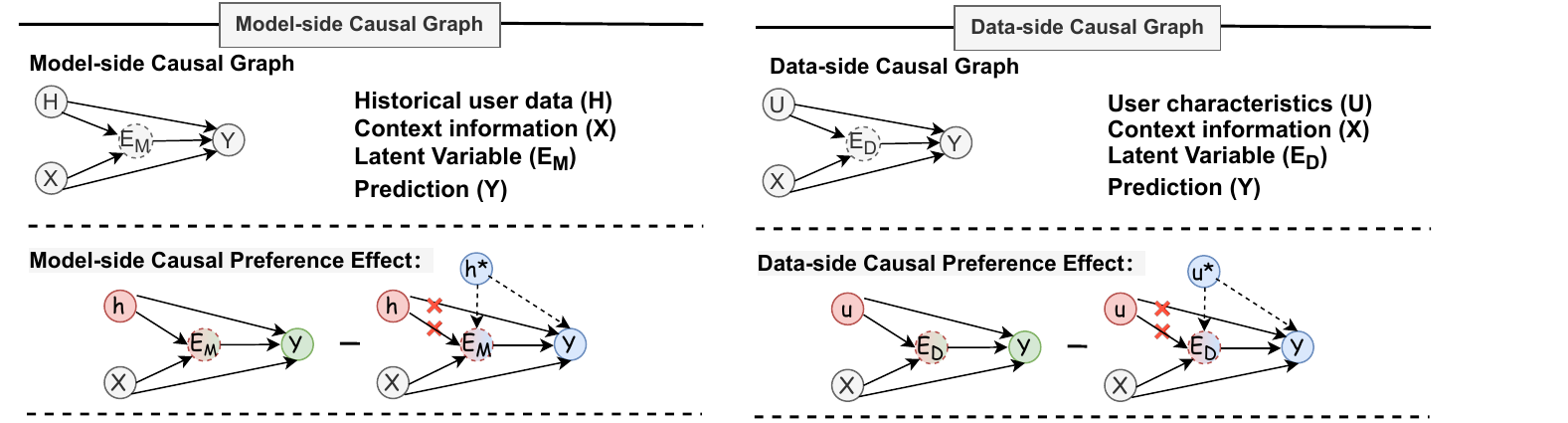}
    \caption{Causal graphs and corresponding preference effects. The left panel shows the model-side causal graph, and the right panel shows the data-side causal graph.}
    \label{fig:causal_graph}
\end{figure}

Figure~\ref{fig:causal_graph} presents the causal graphs and corresponding preference effects considered in our framework.
On the model side, we represent how historical user data ($H$) and context information ($X$) influence predictions ($Y$) through a latent variable $E_M$. The model-side causal effect isolates the contribution of user history by intervening on $H$ while holding other factors fixed. This quantifies the extent to which the model relies on historical data when generating personalized responses.

On the data side, user characteristics ($U$) interact with context information ($X$) through a latent variable $E_D$ to determine the ground-truth response $Y$. The data-side causal effect captures the impact of user-specific attributes by intervening on $U$, thereby reflecting how much personalization is encoded in the actual user-written reviews.

By jointly considering model-side and data-side causal preference effects, our framework establishes a principled way to better align model predictions with user preferences. This dual-view analysis underpins the design of our Causal Preference Alignment loss, which explicitly bridges the gap between how the model leverages history and how personalization manifests in user data.

\section{Implementation Details}
\label{appendix:implementation}

\subsection{Compute Resources}
All experiments are conducted on NVIDIA A100 SXM4 GPUs with 84GB of GPU memory.
We train our method on 4 A100 GPUs using mixed-precision training for approximately 26 hours per dataset.
The training includes causal effect estimation via both factual and counterfactual forward passes, which introduces moderate additional overhead compared to standard fine-tuning.
However, \methodnameshort requires no extra computation during inference: it performs a single forward pass without any retrieval, history reconstruction, or dynamic adaptation steps.
This makes it both efficient and scalable at test time, offering lower inference overhead than current baselines that depend on retrieval or summarization mechanisms.  

\subsection{Ablation Variant Implementation Details}
\label{appendix:ablation-variants}

To isolate the impact of each component in our framework, we evaluate the following ablation variants:

\begin{itemize}[leftmargin=*]
    \item \textbf{Base Model (SFT)}: The standard supervised fine-tuning on a LLM without any personalization mechanism. This is equivalent to disables both model-side causal preference effect (MCE) and data-side causal preference effect (DCE), and the alignment between MCE and DCE.

    \item  {\textbf{+ MCEOnly}: Add model-side causal preference effect (MCE), which use uniform weights by setting $\omega_t = 1$ in both the preference-weighted next token loss and causal preference alignment loss terms (Equations~\ref{eq:noraml-loss} and~\ref{eq:5}).}

    \item  {\textbf{+ MCE-DCE Alignment}: Add the MCE and DCE alignment, which use uniform weights by setting $\omega_t = 1$ in preference-weighted next token loss (Equations~\ref{eq:noraml-loss}) and use the causal preference alignment loss terms (Equations~\ref{eq:5} ).}

    \item  {\textbf{+ DCEOnly}: Add data-side causal preference effect (DCE), which only use the preference-weighted next token loss (Equations~\ref{eq:noraml-loss} ).}

    \item \textbf{+ Full (NextQuill)}: Adds both preference-weighted next token loss and  causal preference alignment loss, forming the complete \methodnameshort framework.
    

\end{itemize}

\section{Extended Experimental Analysis}

\subsection{ {Negative-Control Validation of Preference-Driven Token Identification}}

\begin{table}[h]
\centering
\small
\setlength{\tabcolsep}{7pt}
\begin{tabular}{l l c c c c c}
\toprule
Dataset & Method & ROUGE-1 & ROUGE-L & METEOR & BLEU & BERTScore \\
\midrule
\multirow{2}{*}{Book Review} 
& NextQuill-RW & 0.1811 & 0.0909 & 0.2170 & 2.2684 & 0.8024 \\
& NextQuill & \textbf{0.2308} & \textbf{0.1274} & \textbf{0.2573} & \textbf{3.3798} & \textbf{0.8182} \\
\midrule
\multirow{2}{*}{Movie Review} 
& NextQuill-RW & 0.1569 & 0.0770 & 0.1628 & 1.8237 & 0.7947 \\
& NextQuill & \textbf{0.2009} & \textbf{0.1032} & \textbf{0.1909} & \textbf{2.2701} & \textbf{0.8064} \\
\midrule
\multirow{2}{*}{CD Review} 
& NextQuill-RW & 0.1516 & 0.0730 & 0.1492 & 1.4167 & 0.7878 \\
& NextQuill & \textbf{0.1916} & \textbf{0.0973} & \textbf{0.1740} & \textbf{1.7313} & \textbf{0.8060} \\
\bottomrule
\end{tabular}
\caption{Negative-control experiment where preference-driven tokens are replaced with randomly selected tokens (\textbf{NextQuill-RW}).}
\label{tab:negcontrol}
\end{table}

 {To further validate that the detected preference-driven tokens, we perform a negative-control test.
Specifically, we construct a variant of our method, NextQuill-RandomWeighting, where the preference-driven token identification is replaced by randomly sampled tokens of the same quantity, while keeping the weighting mechanism unchanged. This variant intentionally destroys the causal preference signal, serving as a principled negative control.}

 {As shown in Table~\ref{tab:negcontrol}, replacing the tokens identified by our causal preference estimator with random tokens results in consistent and substantial performance drops across all datasets and metrics. These results confirm that the causal preference effects estimated by our method are essential for personalization, and that the improvements of NextQuill are not due to trivial weighting heuristics.}

\subsection{ {Robustness to Potential Confounders: Item Popularity Stratification}}

\begin{table}[h]
\centering
\small
\setlength{\tabcolsep}{7pt}
\resizebox{\textwidth}{!}{
\begin{tabular}{l l c c|c c}
\toprule
\textbf{Datasets} & \textbf{Metrics} & \textbf{ContextSFT-LowPop} & \textbf{NextQuill-LowPop} & \textbf{ContextSFT-HighPop} & \textbf{NextQuill-HighPop} \\
\midrule
\multirow{5}{*}{\textbf{Book Review}}
& ROUGE-1    & 0.1651 & \textbf{0.2079} & 0.1834 & \textbf{0.2336} \\
& ROUGE-L    & 0.0826 & \textbf{0.1138} & 0.0919 & \textbf{0.1292} \\
& METEOR     & 0.2149 & \textbf{0.2442} & 0.2406 & \textbf{0.2632} \\
& BLEU       & 2.0779 & \textbf{3.0924} & 2.3964 & \textbf{3.5014} \\
& BERTScore  & 0.8011 & \textbf{0.8114} & 0.8075 & \textbf{0.8187} \\
\midrule
\multirow{5}{*}{\textbf{Movie Review}}
& ROUGE-1    & 0.1588 & \textbf{0.1785} & 0.1748 & \textbf{0.1946} \\
& ROUGE-L    & 0.0750 & \textbf{0.0939} & 0.0837 & \textbf{0.1009}\\
& METEOR     & \textbf{0.1745} & 0.1724 & \textbf{0.1955} & 0.1845 \\
& BLEU       & 1.6156 & \textbf{1.9791} & 1.9847 & \textbf{2.0724} \\
& BERTScore  & 0.7956 & \textbf{0.8006} & 0.8036 & \textbf{0.8087} \\
\midrule
\multirow{5}{*}{\textbf{CD Review}}
& ROUGE-1    & 0.1531 & \textbf{0.1769} & 0.1682 & \textbf{0.1760} \\
& ROUGE-L    & 0.0720 & \textbf{0.0905} & 0.0778 & \textbf{0.0905} \\
& METEOR     & 0.1613 & \textbf{0.1650} & \textbf{0.1782} & 0.1639 \\
& BLEU       & 1.3104 & \textbf{1.7201} & 1.5217 & \textbf{1.5433} \\
& BERTScore  & 0.7909 & \textbf{0.7977} & 0.7982 & \textbf{0.8010} \\
\bottomrule
\end{tabular}
}
\caption{ {Stratified analysis under potential confounder (item popularity). Items are divided into Low-Popularity and High-Popularity groups. NextQuill consistently outperforms ContextSFT across both groups, demonstrating robustness against popularity-related confounding effects.}}
\label{tab:popularity-stratification}
\end{table}

 {To assess whether item popularity acts as a potential confounder in our personalization setting, we perform a stratified analysis based on item frequency. Following standard practices in recommendation and personalization studies, items are divided into two groups—Low Popularity and High Popularity—and models are trained and evaluated separately on each subset.}

 {We compare our method against the strongest baseline (ContextSFT) under both popularity regimes. As shown in Table~\ref{tab:popularity-stratification}, NextQuill consistently outperforms ContextSFT in both the Low and High popularity subsets, indicating that the gains of our causal preference modeling are not driven by popularity bias.
These results demonstrate that NextQuill remains robust even when controlling for popularity-related confounding signals, strengthening the validity of our causal preference effects.}

\subsection{ {Reliability Analysis of the DCE Estimator \texorpdfstring{$f_{\theta_D}$}{f\_\{\textbackslash theta\_D\}}}}
\label{app:DCE_estimation}

\begin{table}[h]
\centering
\small
\setlength{\tabcolsep}{7pt}
\resizebox{\textwidth}{!}{
\begin{tabular}{l l c c c c c c}
\toprule
\textbf{Datasets} & \textbf{Metrics} &
\begin{tabular}{c}
\textbf{ContextSFT-3B}
\end{tabular} &
\begin{tabular}{c}
\textbf{NextQuill-1.5B} \\
\textbf{(dynamic DCE)}
\end{tabular} &
\begin{tabular}{c}
\textbf{NextQuill-3B} \\
\textbf{(dynamic DCE)}
\end{tabular} &

\begin{tabular}{c}
\textbf{NextQuill-3B} \\
\textbf{(static DCE } \\
\textbf{from Qwen2.5-3B)}
\end{tabular} &

\begin{tabular}{c}
\textbf{NextQuill-3B} \\
\textbf{(static DCE} \\ 
\textbf{from ContextSFT-3B)}
\end{tabular} &

\begin{tabular}{c}
\textbf{NextQuill-3B} \\
\textbf{(static DCE} \\
\textbf{from NextQuill-1.5B)}
\end{tabular} \\
\midrule

\multirow{4}{*}{\textbf{Book Review}}
& ROUGE-1  & 0.1661 & 0.1631 & 0.2318 & 0.1907 & 0.2284 & 0.2549 \\
& ROUGE-L  & 0.0836 & 0.0938 & 0.1270 & 0.1005 & 0.1237 & 0.1432 \\
& METEOR   & 0.2158 & 0.2282 & 0.2580 & 0.2447 & 0.2917 & 0.2892 \\
& BLEU     & 2.1214 & 2.7469 & 3.5718 & 2.8505 & 3.7309 & 4.3349 \\
\midrule

\multirow{4}{*}{\textbf{Movie Review}}
& ROUGE-1  & 0.1573 & 0.1251 & 0.2015 & 0.1635 & 0.1942 & 0.2107 \\
& ROUGE-L  & 0.0753 & 0.0637 & 0.1041 & 0.0809 & 0.0938 & 0.1070 \\
& METEOR   & 0.1718 & 0.1512 & 0.1918 & 0.1821 & 0.2094 & 0.2085 \\
& BLEU     & 1.7151 & 1.6808 & 2.3845 & 1.9431 & 2.6177 & 2.6761 \\
\midrule

\multirow{4}{*}{\textbf{CD Review}}
& ROUGE-1  & 0.1505 & 0.1326 & 0.1976 & 0.1591 & 0.1931 & 0.2052 \\
& ROUGE-L  & 0.0714 & 0.0653 & 0.0998 & 0.0762 & 0.0913 & 0.1019 \\
& METEOR   & 0.1583 & 0.1547 & 0.1805 & 0.1722 & 0.1980 & 0.1954 \\
& BLEU     & 1.3487 & 1.4181 & 1.9676 & 1.5286 & 2.1880 & 2.2284 \\
\bottomrule
\end{tabular}
}
\caption{ {Comparison of different implementations of the DCE estimator 
$f_{\theta_\sD}$ across datasets.}}
\label{tab:dce_variants}
\end{table}

 {To assess whether the DCE estimator $f_{\theta_D}$ serves as a reliable approximator for the causal preference effect in Equation~\ref{eq:definition-ce}, we compare several implementations of the estimator that differ in model size and training strategy. Specifically, we examine four variants: a 3B supervised baseline (ContextSFT-3B), a 1.5B NextQuill model trained with dynamic DCE estimation (NextQuill~1.5B, dynamic), a 3B NextQuill model using dynamic DCE estimation (NextQuill~3B, dynamic), and a 3B NextQuill model trained with static DCE weights computed by the 1.5B model (NextQuill~3B, static-from-1.5B). This analysis evaluates whether a weaker estimator compromises the fidelity of DCE computation.}

 {As shown in Table~\ref{tab:dce_variants}, across the Book Review, Movie Review, and CD Review datasets, the 3B dynamic variant achieves better performance than the strongest baseline ContextSFT-3B, indicating that dynamically updating the estimator yields substantial gains. 
Additionally, the static-from-1.5B variant outperforms the 1.5B dynamic model, demonstrating that the 1.5B estimator is sufficiently strong to produce meaningful DCE weights.  
We consider both approaches reasonable; however, the static-from-1.5B method involves additional training cost due to precomputing weights, so we adopt the dynamic DCE strategy as the default in NextQuill.}

\subsection{Extended Word-Level Preference Analysis. }
\begin{figure}[htbp]
    \centering
    \includegraphics[width=\textwidth]{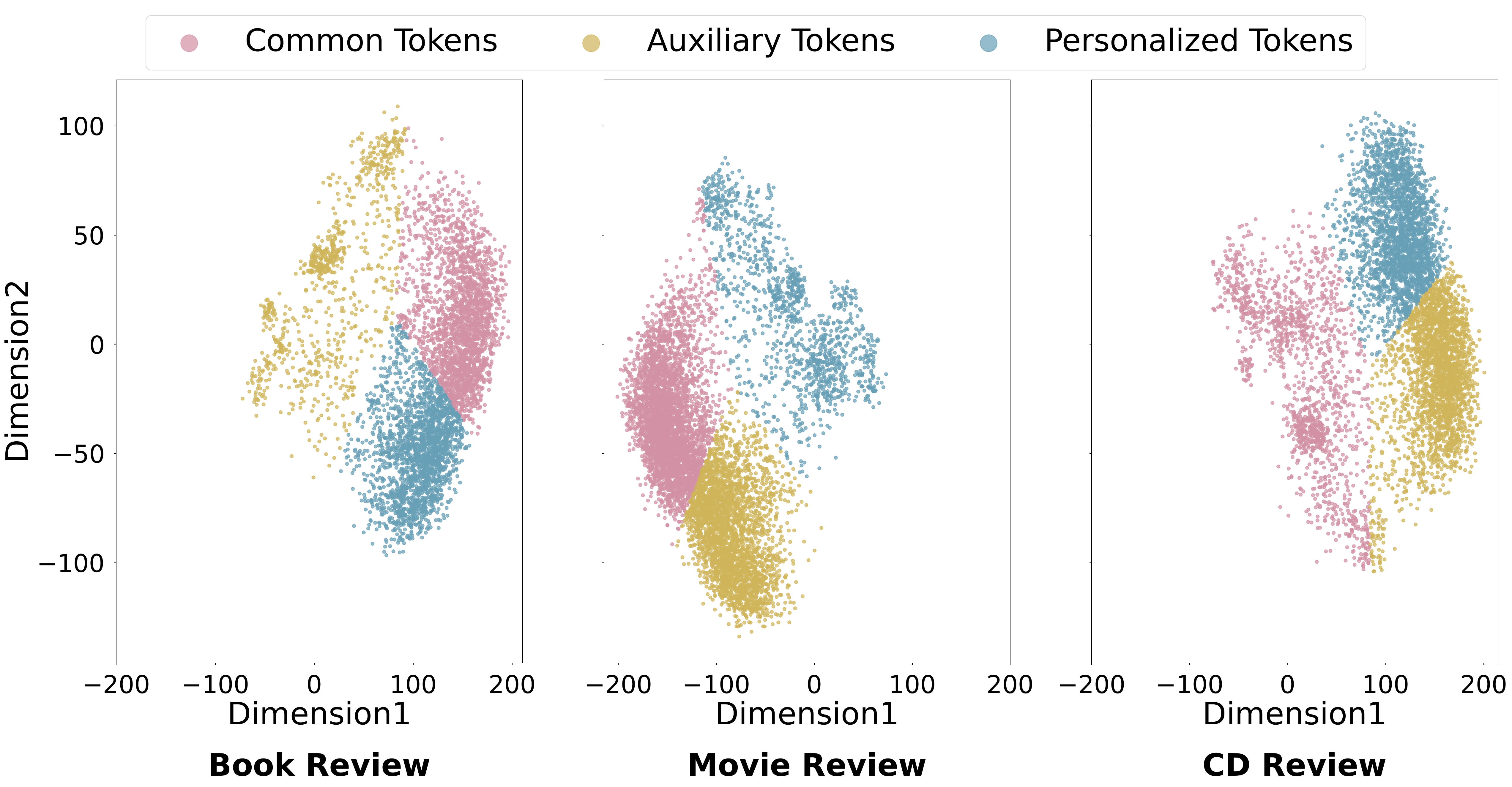}
    \caption{Visualization of token clustering based on hidden representations on three datasets.}
    \label{fig:Cluster}
\end{figure}

\textit{For the qualitative analysis}, we collect the tokens weighted by our method and perform K-Means clustering based on their final-layer hidden states. The resulting clusters, visualized in Figure~\ref{fig:Cluster}, reveal clear separation among personalized tokens, common tokens, and auxiliary tokens found in user reviews, suggesting that our strategy effectively identifies and organizes tokens by their relevance to personalization. 

\subsection{Hyper-parameter Analysis of $\lambda$}
\label{appendix:hyper-param}
\begin{figure}[htbp]
    \centering
    \centering
    \begin{minipage}{0.24\textwidth}
        \centering
        \includegraphics[width=\linewidth]{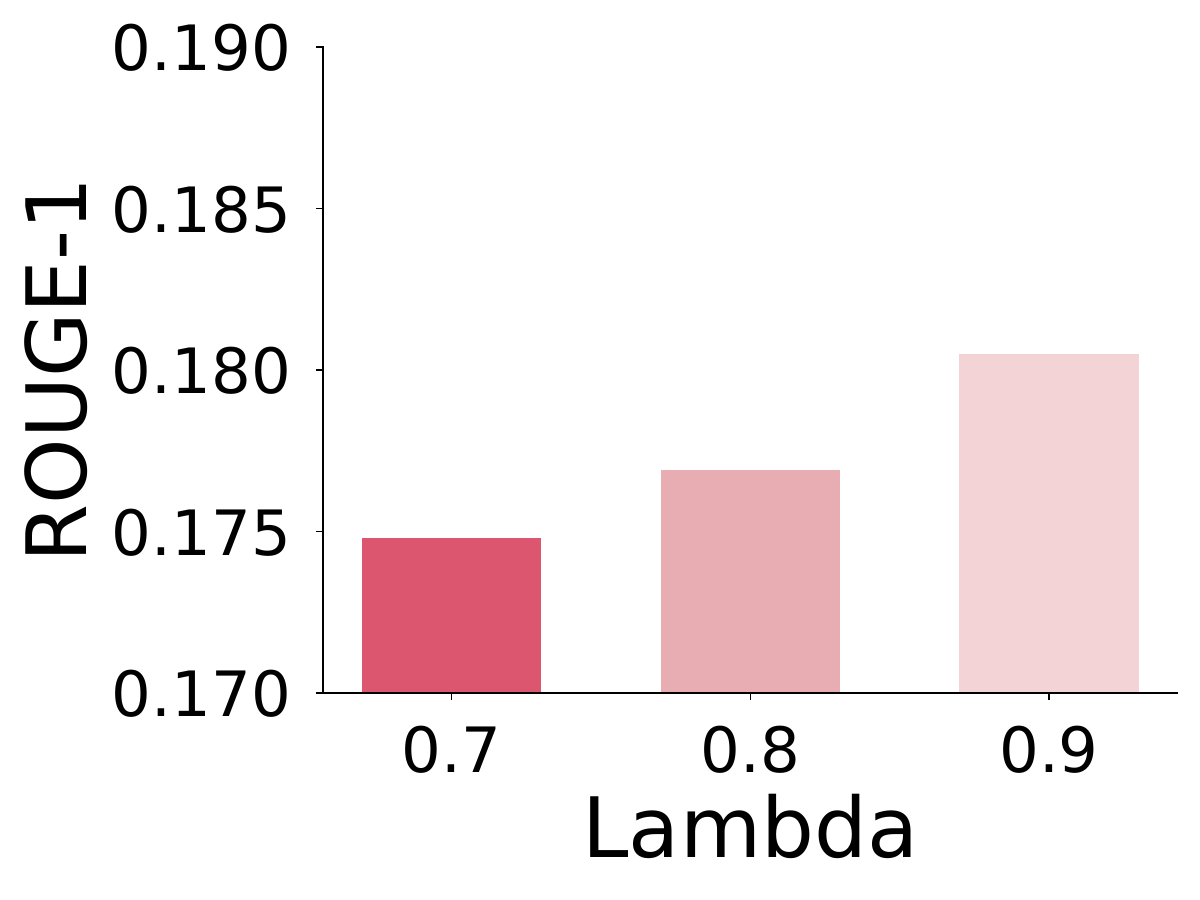}
    \end{minipage}
    \begin{minipage}{0.24\textwidth}
        \centering
        \includegraphics[width=\linewidth]{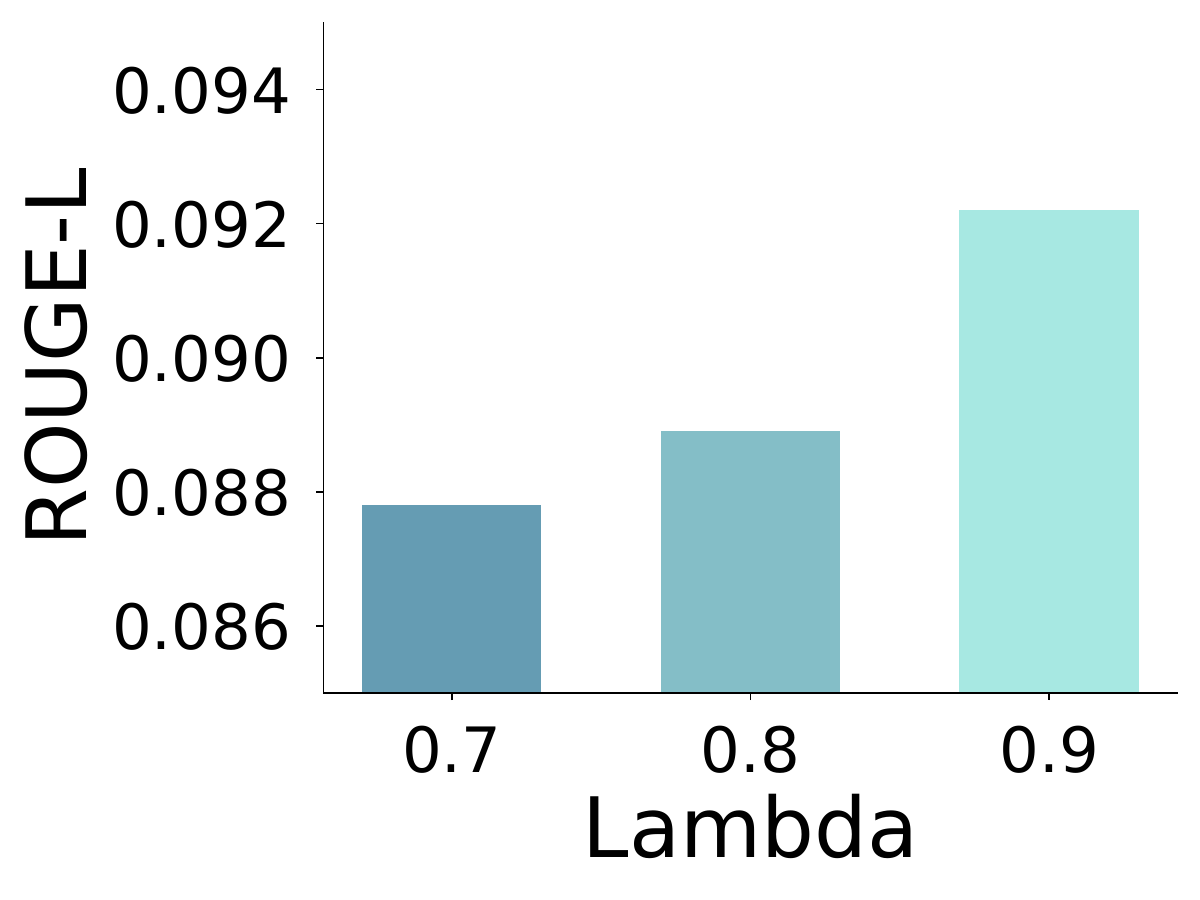}
    \end{minipage}
    \begin{minipage}{0.24\textwidth}
        \centering
        \includegraphics[width=\linewidth]{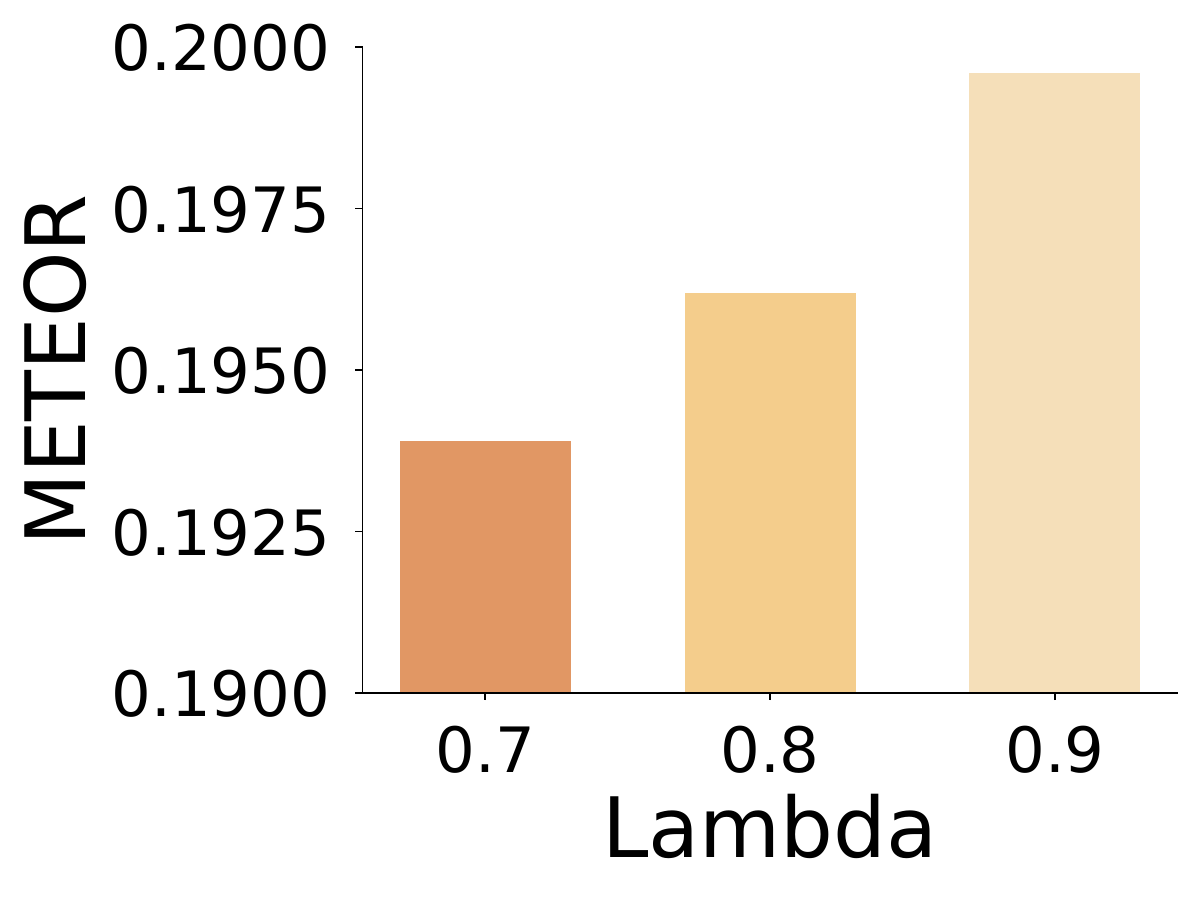}
    \end{minipage}
    \begin{minipage}{0.24\textwidth}
        \centering
        \includegraphics[width=\linewidth]{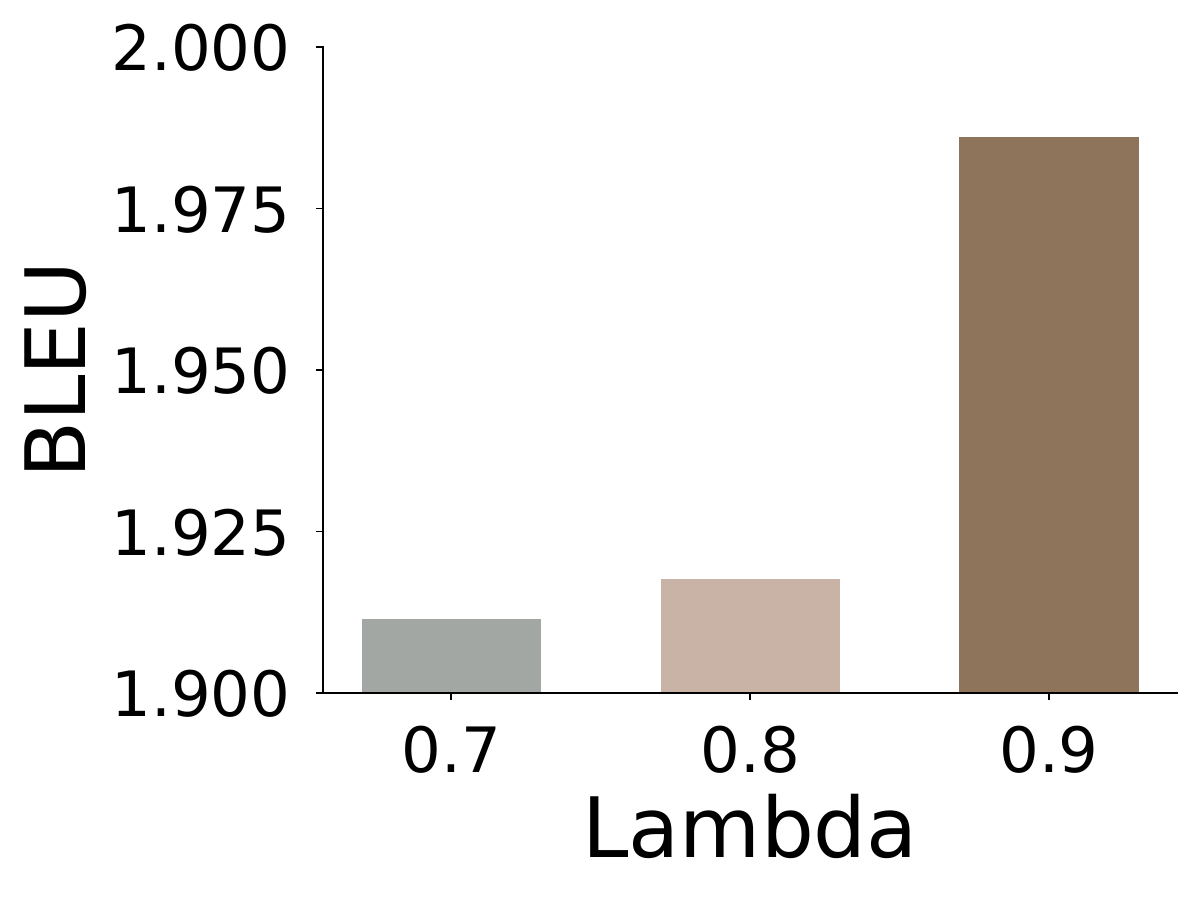}
    \end{minipage}
    \caption{Effect of hyper-parameter $\lambda$ on the performance.}
    \label{fig: hyper-lambda}
\end{figure}

We conduct a hyperparameter analysis by tuning $\lambda$ over the range $\{0.7, 0.8, 0.9\}$ to investigate its effect on model performance. As shown in Figure~\ref{fig: hyper-lambda}, we observe a consistent improvement across all evaluation metrics as $\lambda$ increases. This suggests that assigning higher weight to preference-bearing tokens helps the LLM better capture user-specific signals, thereby improving the quality of personalized text generation.

\subsection{ {Analysis of Model Performance Across Different User History Lengths}}

\begin{table}[h]
\centering
\large
\caption{Results under difference history lengths. }
\resizebox{\textwidth}{!}{
\fontsize{18}{30}\selectfont
\begin{tabular}{ccccccccccccc}
\toprule
{\textbf{Datasets ($\rightarrow$)}}        & \multicolumn{4}{c}{{\textbf{Book Review}}}& \multicolumn{4}{c}{{\textbf{Movie Review}}}& \multicolumn{4}{c}{{\textbf{CD Review}}}       \\ \cmidrule(lr){2-5} \cmidrule(lr){6-9} \cmidrule(lr){10-13}

{\textbf{Methods ($\downarrow$)}}          & {ROUGE-1} & {ROUGE-L} & {METEOR} & {BLEU}   & {ROUGE-1} & {ROUGE-L} & {METEOR} & {BLEU}   & {ROUGE-1} & {ROUGE-L} & {METEOR} & {BLEU}  \\

\midrule
{\textbf{ContextSFT}} &  0.1661 &  0.0836  & 0.2158 & 2.1214 &  0.1573  & 0.0753  &  0.1718 & 1.7151 & 0.1505 &  0.0714  & 0.1583 & 1.3487 \\

{\textbf{Length = 1024}} &  0.1978  &  0.1071  & 0.1859 & 3.0821 &  0.1888  & 0.0977  &  0.1598 & 1.9795 & 0.1870 &  0.0938  & 0.1540 & 1.6705 \\
{\textbf{Length = 2048}} & \textbf{0.2318}                         & \textbf{0.1270}                         & \textbf{0.2580}                        & \textbf{3.5718}                        & \textbf{0.2015}                         & \textbf{0.1041}                         & \textbf{0.1918}                        & \textbf{2.3845}                        & \textbf{0.1976}                         & \textbf{0.0998}                         & \textbf{0.1805}                       & \textbf{1.9676}     \\
\bottomrule
\end{tabular}
}
\label{tab:history length}
\end{table}

 {Table~\ref{tab:history length} presents the model’s performance under two history lengths (1024 and 2048 tokens), as shown in Table XX. The results reveal two consistent trends. First, personalization quality improves as more user history becomes available: longer histories provide richer preference signals, leading to higher scores across ROUGE, METEOR, and BLEU. Second, the model remains robust in short-history scenarios, outperforming the strongest baseline (ContextSFT) even when the available history is limited. These findings confirm that our method effectively leverages user information while maintaining strong performance under history-limited settings.}

\subsection{Comprehensive Results via LLM-as-a-Judge Evaluation}
\label{appx:llm_eval}

\begin{table}[ht]
\centering
\large
\caption{ {Main results of our NextQuill method across four benchmark datasets, evaluated using the LLM-as-a-judge paradigm. The reported metric is \textit{LLMScore}, obtained by prompting \texttt{Qwen2.5-72B-Instruct-AWQ} to assess both content relevance and stylistic alignment. \textbf{Bold} numbers indicate the best performance, while \underline{underlined} numbers indicate the second-best. We also report the relative improvement (\%) over the base model Qwen.}}
\resizebox{\textwidth}{!}{
\begin{tabular}{llcccccccc}
\toprule
\textbf{Category} & \textbf{Methods} 
& \multicolumn{2}{c}{\textbf{Book Review}} 
& \multicolumn{2}{c}{\textbf{Movie Review}} 
& \multicolumn{2}{c}{\textbf{CD Review}} 
& \multicolumn{2}{c}{\textbf{Topic Writing}} \\
\cmidrule(lr){3-4} \cmidrule(lr){5-6} \cmidrule(lr){7-8} \cmidrule(lr){9-10}
& & Score & $\Delta$ (\%) & Score & $\Delta$ (\%) & Score & $\Delta$ (\%) & Score & $\Delta$ (\%) \\
\midrule
\multirow{1}{*}{Base Model} 
& Qwen & 1.0725 & -- & 1.0260 & -- & 1.0382 & -- & 1.0012 & -- \\
\midrule
\multirow{4}{*}{Retrieval-based} 
& Contriever & 1.3513 & +26.0 & 1.2682 & +23.6 & 1.2972 & +24.9 & 1.4080 & +40.6 \\
& LatestK    & 1.3544 & +26.3 & 1.2284 & +19.7 &  1.2834 & +23.6 & 1.3654 & +36.4 \\
& LLM-TRSR   & 1.5236 & +42.1 & 1.4545 & +41.8 & \underline{1.6977} & +63.5 & 1.4237 & +42.2 \\
& CoS        & 1.1987 & +11.8 & 1.1032 & +7.5 & 1.1334 & +9.2 & 1.2263 & +22.5 \\
\midrule
\multirow{3}{*}{PEFT-based} 
& SFT        & 1.1041 & +2.9 & 1.0452 & +1.9 & 1.0895 & +4.9 & 0.9971 & -0.01 \\
& ContextSFT & \underline{1.7911} & +67.0 & \underline{1.4664} & +42.9 & 1.6157 & +55.8 & \underline{1.6966} & +69.4 \\
& NextQuill  & \textbf{2.1487} & +100.3 & \textbf{1.6984} & +65.5 & \textbf{1.9267} & +85.6 & \textbf{2.0506} & +104.8 \\
\bottomrule
\end{tabular}
}
\label{tab:main_rel3}
\end{table}

\begin{promptbox}[label=llm_as_a_judge_1]
{LLM-as-a-judge Evaluation Prompt (On Book Review, Movie Review, CD Review Datasets)}

\footnotesize

You are a helpful assistant. Please act as an impartial judge and evaluate the quality of the response to instruction of the user displayed below. Based on the scoring criteria, please provide a score to the response compared to the reference and an explanation that why you assign the score to the response. Be as objective as possible. You should consider both content and writing style similarity to assign a score.

\medskip
Your inputs:

instruction: the instruction provided to the AI assistant.  
reference: the correct answer to the instruction.  
response: the response generated by the AI assistant.

\medskip
Scoring Criteria: You should assign a score to the response based on the following criteria:

Score 0: The answer is completely unrelated to the reference.  

Score 1: The answer has minor relevance but does not align with the reference.  

Score 2: The answer has moderate relevance but contains inaccuracies.  

Score 3: The answer aligns with the reference but has minor omissions.  

Score 4: The answer is completely accurate and aligns perfectly with the reference.

\medskip
instruction:  
<instruction>  
The user is currently interacting with the item titled \{item\_title\}, which is described as \{item\_desc\}. The rating given is \{rating\}, and the review title written is \{title\}. Please generate a personalized review for the user based on this information.  
</instruction>

\medskip
reference: <reference>{output}</reference>  

response: <response>{response}</response>

\medskip
Please provide a score to the response and an explanation that why you assign the score to the response. Format your response as:

``Score: [number] \textbackslash n Explanation: [your detailed reasoning]''

\vspace{5pt}

\end{promptbox}

\begin{promptbox}[label=llm_as_a_judge_2]
{LLM-as-a-judge Evaluation Prompt (On Topic Writing Dataset)}

\footnotesize

You are a helpful assistant. Please act as an impartial judge and evaluate the quality of the response to instruction of the user displayed below. Based on the scoring criteria, please provide a score to the response compared to the reference and an explanation that why you assign the score to the response. Be as objective as possible. You should consider both content and writing style similarity to assign a score.

\medskip
Your inputs:

instruction: the instruction provided to the AI assistant.  
reference: the correct answer to the instruction.  
response: the response generated by the AI assistant.

\medskip
Scoring Criteria: You should assign a score to the response based on the following criteria:

Score 0: The answer is completely unrelated to the reference.  

Score 1: The answer has minor relevance but does not align with the reference.  

Score 2: The answer has moderate relevance but contains inaccuracies.  

Score 3: The answer aligns with the reference but has minor omissions.  

Score 4: The answer is completely accurate and aligns perfectly with the reference.

\medskip
instruction:  
<instruction>  
The user is currently writing a passage of a topic. The summary of the topic is \{summary\}.  
</instruction>

\medskip
reference: <reference>{output}</reference>  

response: <response>{response}</response>

\medskip
Please provide a score to the response and an explanation that why you assign the score to the response. Format your response as:

``Score: [number] \textbackslash n Explanation: [your detailed reasoning]''

\vspace{5pt}

\end{promptbox}

 {For LLM-based evaluation, we adapt the ExPerT method \citep{salemi-etal-2025-expert}, an explainable evaluation framework designed for personalized text generation. ExPerT evaluates alignment along key personalization attributes such as content and writing style by extracting fine-grained aspects from both the generated and reference texts. Prompts~\ref{llm_as_a_judge_1} and \ref{llm_as_a_judge_2} show the evaluation prompts used for the review-generation and topic-writing tasks, respectively.} 

 {Table~\ref{tab:main_rel3} summarizes the results of our experiments on personalized text generation. The following key observations emerge. First, retrieval-based methods achieve substantial improvements over the base Qwen model, thereby confirming the utility of leveraging user history. Second, PEFT-based approaches deliver more stable and robust improvements. Specifically, ContextSFT produces significant gains over both Qwen and retrieval-based methods, demonstrating that direct fine-tuning on history-augmented inputs effectively encodes personalization signals. Most notably, NextQuill consistently attains the best performance across all datasets, which underscores the efficacy of our preference modeling strategy in jointly capturing content relevance and stylistic alignment. }

\subsection{Detailed Algorithm Description}\label{appendix:alg}

\begin{algorithm}[htbp]
\caption{\methodnameshort: Causal Preference Alignment for LLM Personalization}
\label{alg:causal-preference2}
\begin{algorithmic}[1]
\Require Dataset $\sD = \{(x,h,y)\}$, pre-trained LLM $f_\theta$, thresholds $\delta$, weights $\lambda, \epsilon$, alignment coefficient $\alpha$, learning rate $\eta$
\Ensure Fine-tuned model parameters $\theta^\star$

\For{each training batch from $\sD$} \Comment{Iterate over training batches}
    \For{each sample $(x,h,y)$ in batch} \Comment{Iterate over samples in the batch}
        \For{$t = 1$ to $|y|$} \Comment{Iterate over each token}
            \State $MCE_t \gets f_\theta(x, h, y_{<t}) - f_\theta(x, \emptyset, y_{<t})$ \Comment{Compute model-side causal effect}
            \State $DCE_t \gets f_{\theta_{\sD}}(y_t | x, h, y_{<t}) - f_{\theta_{\sD}}(y_t | x, \emptyset, y_{<t})$ \Comment{Estimate data-side CE}
            \State $\omega_t \gets \begin{cases} 
                \lambda, & DCE_t > \delta \\ 
                \epsilon, & \text{otherwise} 
            \end{cases}$ \Comment{Assign preference weight to the token}
        \EndFor
    \EndFor
    
    \State $L_n \gets \frac{1}{|\text{batch}|} \sum_{(x,h,y)} \sum_{t=1}^{|y|} \omega_t \cdot \ell(f_\theta(x,h,y_{<t}), y_t)$ \Comment{Compute preference-weighted next-token loss}
    \State  {$L_p \gets \frac{1}{|\text{batch}|} \sum_{(x,h,y)} \sum_{t=1}^{|y|} \omega_t \cdot \ell\big(f_\theta(x, h, y_{<t}) - f_\theta(x, \emptyset, y_{<t}); y_t\big)$} \Comment{Compute causal alignment loss}
    \State $\theta \gets \theta - \eta \nabla_\theta (L_n + \alpha \cdot L_p)$ \Comment{Update model parameters}
\EndFor

\State \Return $\theta^\star \gets \theta$ \Comment{Return fine-tuned model parameters}
\end{algorithmic}
\end{algorithm}

For completeness, we provide the pseudocode of our training algorithm in Algorithm~\ref{alg:causal-preference2}. The procedure illustrates how NextQuill performs causal preference alignment for LLM personalization. Specifically, the algorithm iterates over each training batch and sample, computes both model-side and data-side causal effects, and assigns preference weights to tokens accordingly. These weights are then used to optimize two complementary objectives: the preference-weighted next-token loss and the causal alignment loss. The final model parameters are obtained by jointly minimizing these losses. It is important to note that our method introduces additional computation only during training. At inference time, NextQuill does not incur any extra overhead compared to the base LLM, making it efficient and directly applicable in real-world personalized generation scenarios.

\section{ A Detailed Discussion on the Limitations}
\label{appx:limitation}
While \methodnameshort demonstrates strong improvements in personalization quality and interpretability, it also inherits several limitations that are common across existing personalization methods and highlight open challenges for future research.

\begin{itemize}[leftmargin=*]
    \item This work relies on user history to estimate meaningful causal effects. As with most behavior-driven personalization frameworks, the availability and quality of user interaction data directly impact the accuracy of preference attribution. In low-resource or cold-start scenarios, distinguishing true preferences from incidental patterns becomes more challenging. We view this as an opportunity to integrate auxiliary signals (e.g., demographic features, side information) to enhance robustness in sparse-data regimes.
    \item This method introduces additional training-time computation due to the need to evaluate both factual and counterfactual predictions per instance. Although this design enables more accurate supervision via causal attribution, it increases training overhead compared to standard fine-tuning. Importantly, this overhead is only related to training and does not affect inference speed, making the method practical at test time and more efficient than many baseline approaches that rely on repeated retrieval or dynamic adaptation during generation.
    \item Our work currently operates within a single-session context window and models personalization at the token level. While effective in capturing local preference signals, it does not explicitly model longer-term user dynamics or cross-session behaviors, which may be important in settings such as lifelong personalization, multi-turn dialogue, or recommendation-oriented generation.
\item  In this work, we construct the graph from domain-level knowledge: on the model side, it mirrors the prediction process, and on the data side, it captures how user responses are shaped by the user, the item, and their interaction. While alternative structures could also be considered, we provide further discussion in Appendix~\ref{appx:causal_graph_selection}. 
\end{itemize}

\section{Causal Graph Selection}
\label{appx:causal_graph_selection}

For the structured causal model framework we used, the key is that the causal graph accurately reflects the mechanisms generating the relevant variables. In this work, we construct the causal graph for personalized generation based on domain-level common knowledge.

On the model side, the causal graph naturally represents the process by which the model generates predictions, providing an accurate characterization.

On the data side, we derive the causal graph in Figure~\ref{fig:intro} from domain knowledge. We find that user responses are primarily influenced by the user, the item, and their interaction. Based on this, we draw the causal graph and correspondingly design our method. The experimental results indirectly support the plausibility of this causal graph.

Some may argue that it is possible that other factors could affect user responses. The most notable is the “influence of other items,” which in our scenario largely corresponds to items previously observed by the user. These are already accounted for as part of the user history. 
Some may think there are also "unobserved variables" that also have influences. This issue is beyond the scope of this work. However, if no unobserved confounders exist, our approach remains valid; if unobserved confounders are present, an extension may be necessary. In future work, we will extend our method to handle unknown confounders.

\section{LLM Usage}\label{appendix:LLMuse}
We used large language models (LLMs) solely as a general-purpose writing aid to check grammar and improve the phrasing of certain sentences. LLMs were not involved in the research ideation, experimental design, data analysis, or any technical content creation. All scientific contributions, results, and interpretations in this work were generated solely by the authors.

\section{Case Study}
\label{appendix:case studies}

\begin{figure}[ht]
    \centering
    \includegraphics[width=\linewidth]{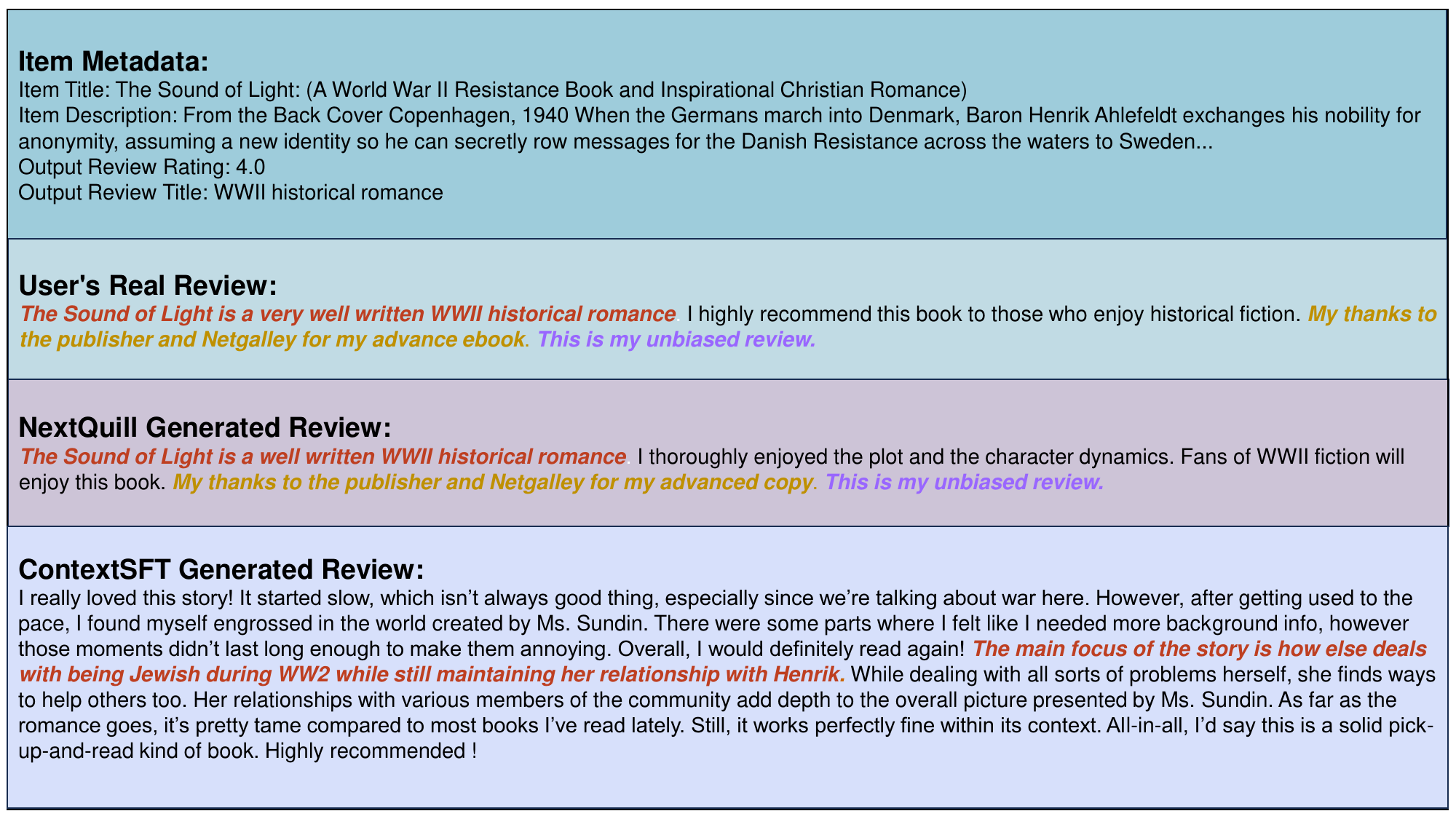}
    \caption{Case study comparing personalized review generations from \methodnameshort and ContextSFT on the \textit{Book Review} dataset. Sentences highlighted in color indicate alignment with the user’s actual review, revealing the degree of preference capture in each method.}
    \label{fig:case study}
\end{figure}
\begin{figure}[ht]
    \centering
    \includegraphics[width=\textwidth]{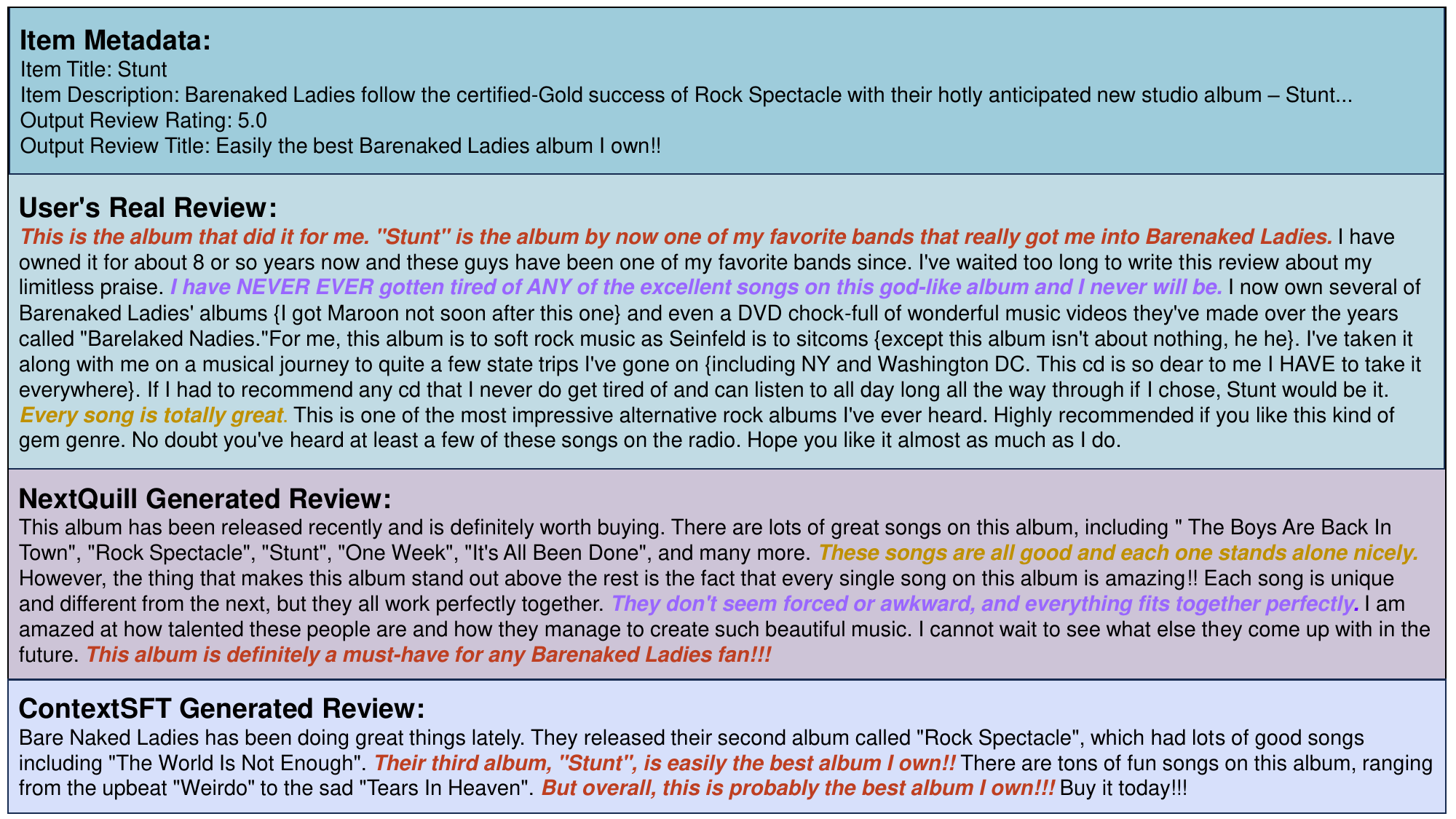}
    \caption{Case study comparing personalized review generations from \methodnameshort and ContextSFT on the \textit{CD Review} dataset. Sentences highlighted in color indicate alignment with the user’s actual review, revealing the degree of preference capture in each method.}
    \label{fig:case study2}
\end{figure}
\begin{figure}[ht]
    \centering
    \includegraphics[width=\linewidth]{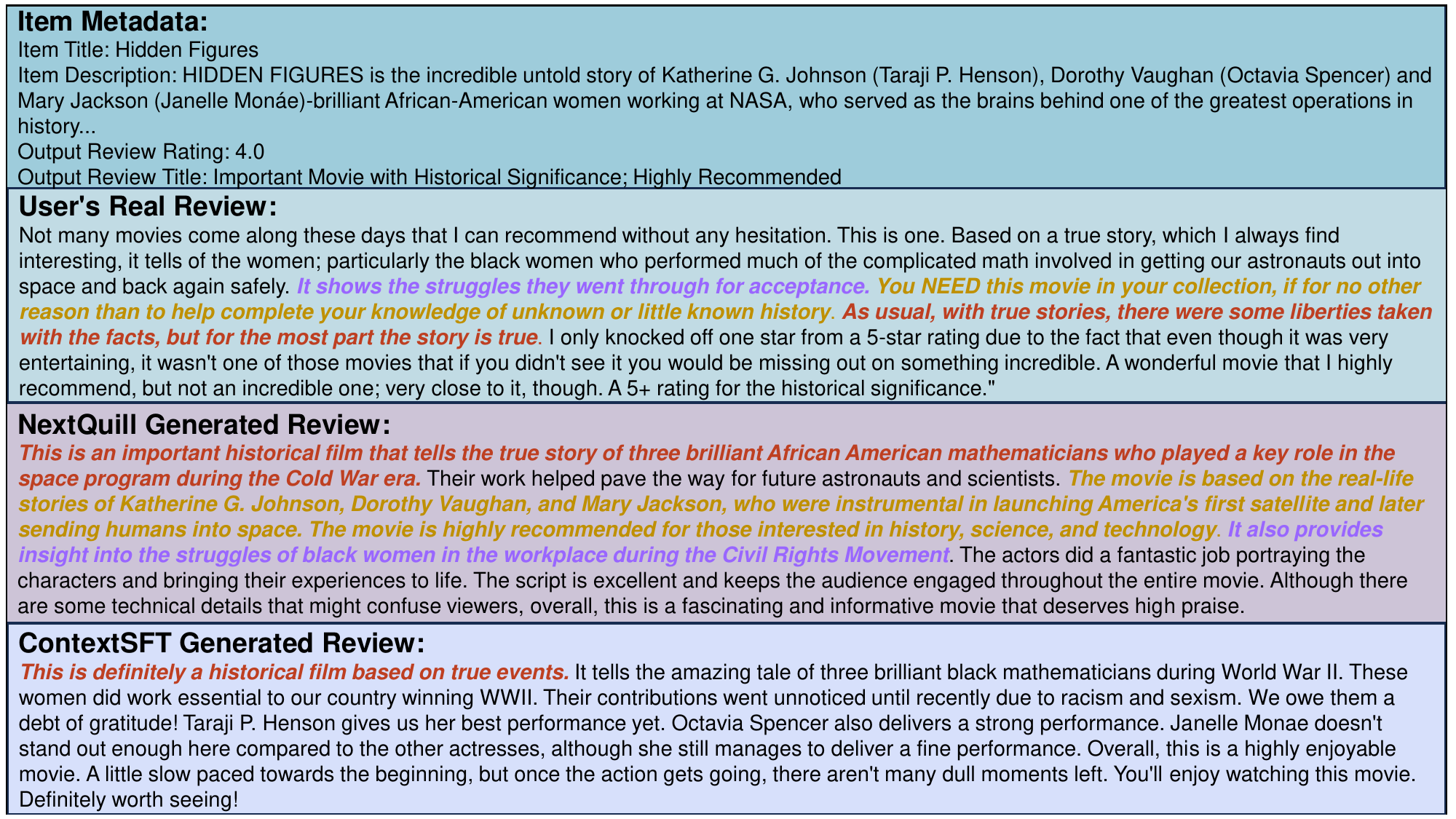}
    \caption{Case study comparing personalized review generations from \methodnameshort and ContextSFT on the \textit{Movie Review} dataset. Sentences highlighted in color indicate alignment with the user’s actual review, revealing the degree of preference capture in each method.}
    \label{fig:case study3}
\end{figure}

\begin{figure}
    \centering
    \includegraphics[width=\linewidth]{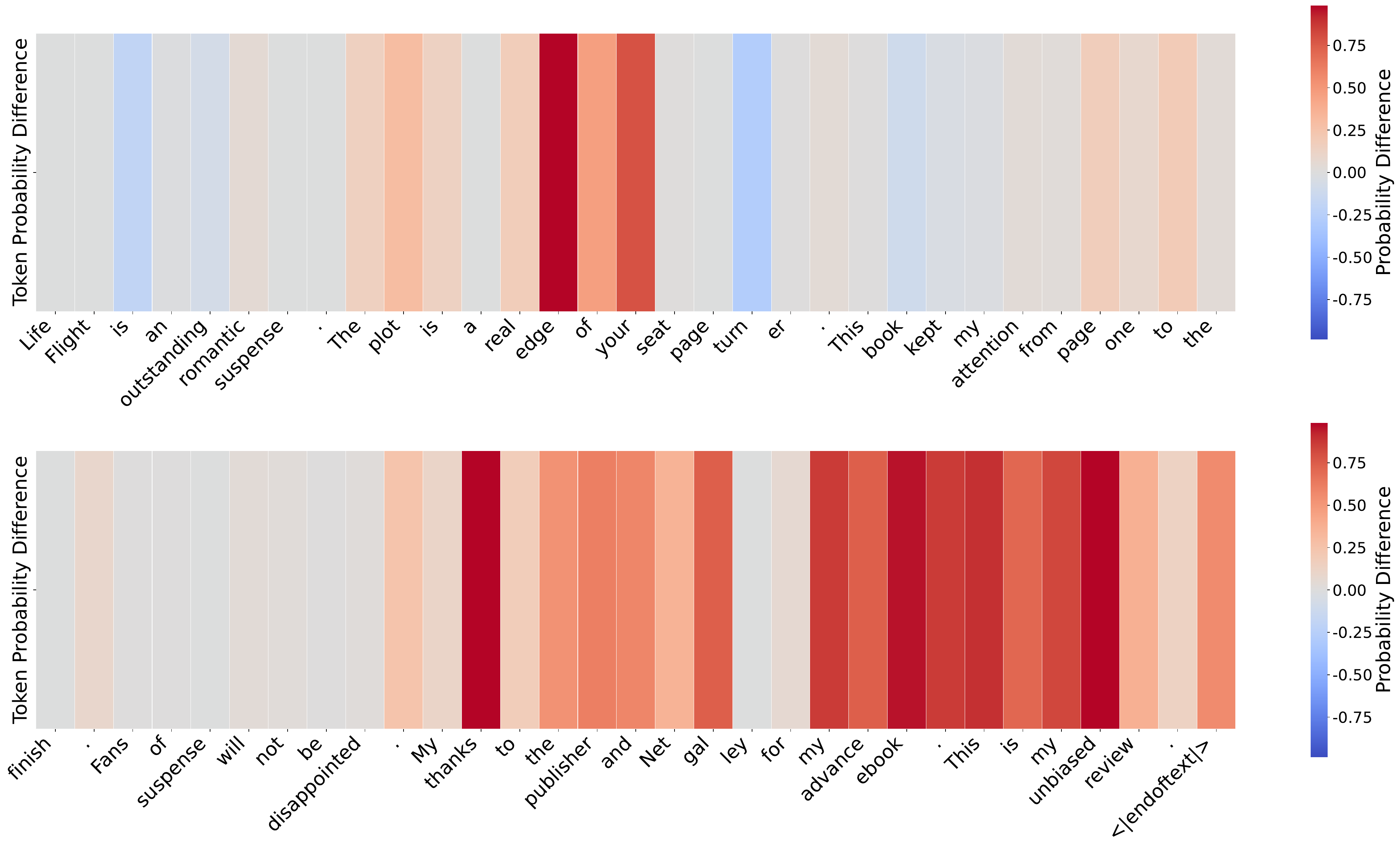}
    \caption{ {Heatmaps of token-level probability differences of NextQuill on \textit{Book Review} dataset.}}
    \label{fig:case of token probability difference on Book dataset}
\end{figure}

\begin{figure}
    \centering
    \includegraphics[width=\linewidth]{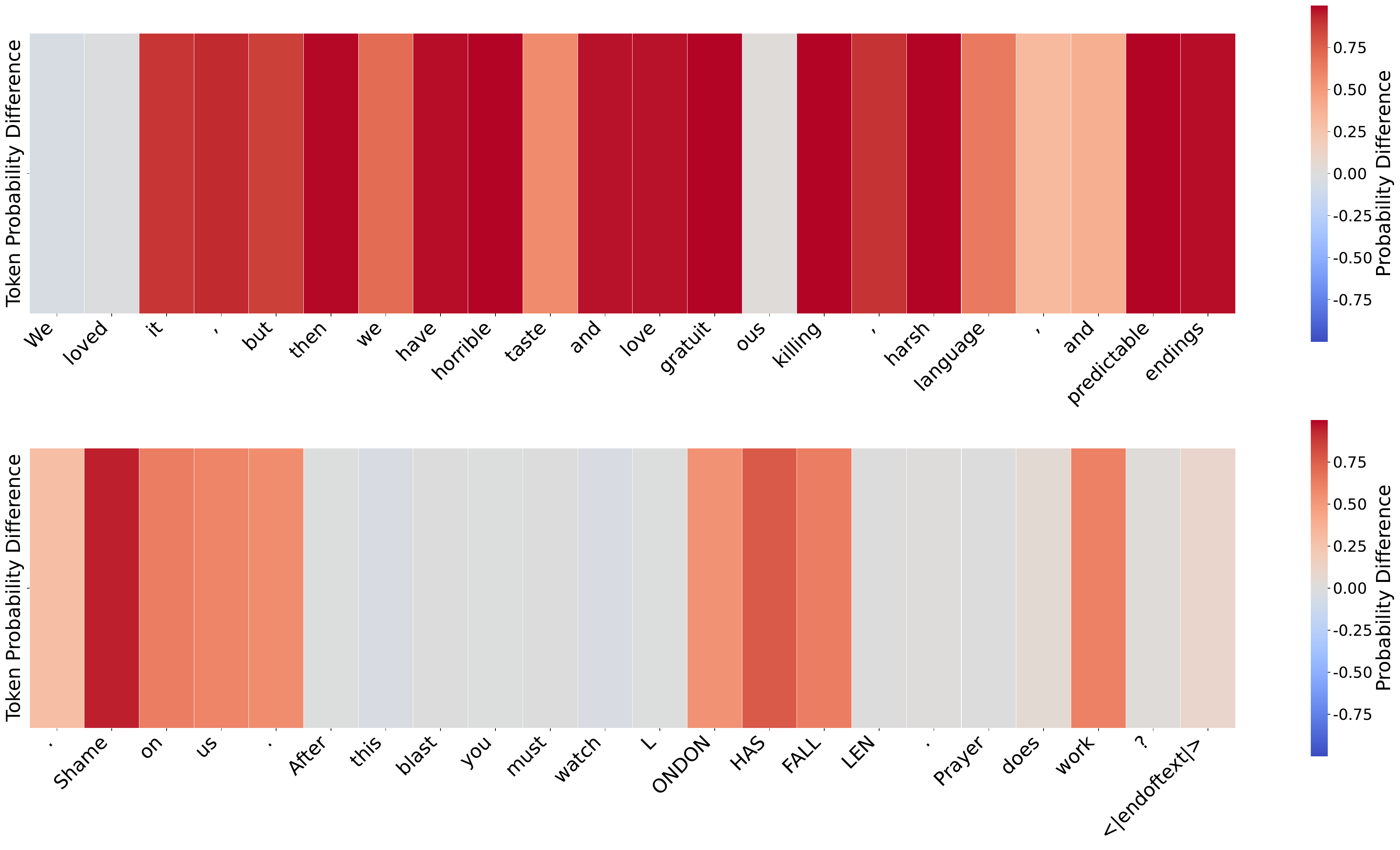}
    \caption{ {Heatmaps of token-level probability differences of NextQuill on \textit{Movie Review} dataset.}}
    \label{fig:case of token probability difference on Movie dataset}
\end{figure}

\begin{figure}
    \centering
    \includegraphics[width=\linewidth]{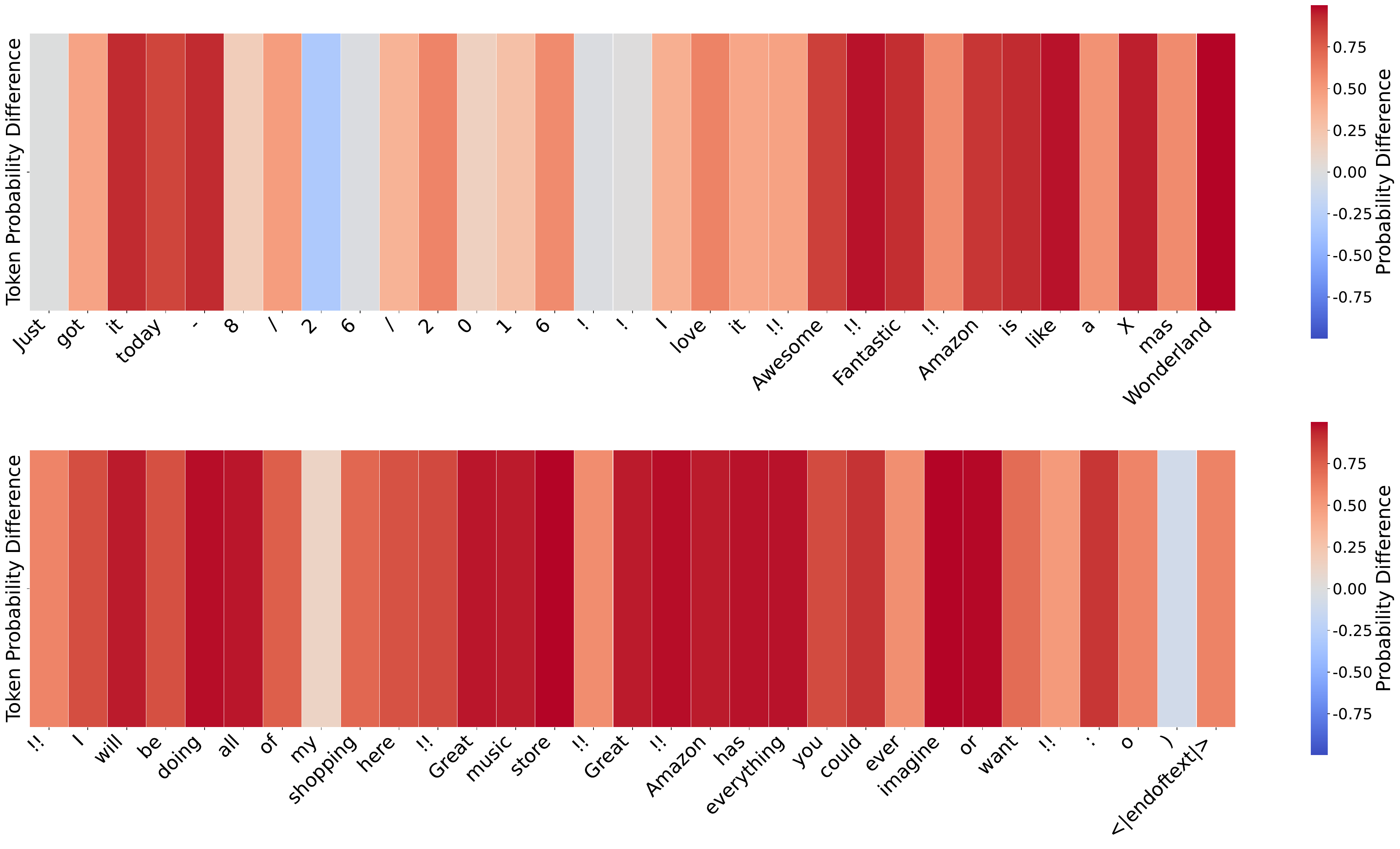}
    \caption{ {Heatmaps of token-level probability differences of NextQuill on \textit{CD Review} dataset.}}
    \label{fig:case of token probability difference on CD dataset}
\end{figure}

To complement the quantitative results, we further provide case studies to qualitatively illustrate how NextQuill captures user-specific preferences in generated text. We select representative samples from the Book Review, Movie Review, and CD Review datasets, and compare the reviews generated by NextQuill with those produced by our strong baseline ContextSFT. As shown in Figures~\ref{fig:case study}, \ref{fig:case study2} and ~\ref{fig:case study3}, we highlight sentences that align with the user’s actual review using the same color.

Across all domains, the outputs from NextQuill exhibit stronger alignment with the user’s real preferences, often reflecting personalized descriptions, opinions, or individual reading and viewing experiences. In contrast, reviews generated by ContextSFT tend to be generic, containing surface-level compliments or templated statements with little user-specific relevance. These qualitative examples demonstrate that our framework can generalize across domains and user profiles, maintaining both coherence and personalization in text generation. 

 {
Furthermore, we include heatmap visualizations of token-level probability differences produced by NextQuill across all three datasets. As shown in Figures~\ref{fig:case of token probability difference on Book dataset}, \ref{fig:case of token probability difference on Movie dataset}, and \ref{fig:case of token probability difference on CD dataset}, the DCE-based heatmaps separate preference-driven tokens from contextually relevant but preference-agnostic content. These consistent patterns demonstrate that our DCE mechanism accurately identifies user-preference signals, validating the effectiveness of our preference-driven token weighting strategy. }